\documentclass[prd,preprint,superscriptaddress,preprintnumbers,eqsecnum,showpacs,nofootinbib,nobibnotes,noeprint]{revtex4-1}
\usepackage[linkcolor=blue,citecolor=blue,urlcolor=blue,colorlinks=true,breaklinks]{hyperref} 
\usepackage{amsfonts,bm}
\usepackage{amsfonts,amssymb,amsmath}
\usepackage{bbold}
\usepackage{graphicx}
\usepackage{enumitem}
\usepackage[english]{babel}
\usepackage{slashed}
\usepackage[usenames]{xcolor}
\usepackage{mathtools}
\usepackage[makeroom]{cancel}
\usepackage{color}

\def\srm#1{{\rm{\scriptscriptstyle #1}}}

\newcommand{\be}{\begin{equation}}
\newcommand{\bea}{\begin{eqnarray}}
\newcommand{\ee}{\end{equation}}
\newcommand{\eea}{\end{eqnarray}}

\def\1eq#1{Eq.~(\ref{#1})}

\def\2eqs#1#2{Eqs.~(\ref{#1}) and~(\ref{#2})}
\def\3eqs#1#2#3{Eqs.~(\ref{#1}),~(\ref{#2}) and~(\ref{#3})}

\def\fig#1{Fig.~\ref{#1}}

\def\ie{{\it i.e.}, }
\def\eg{{\it e.g.}, }

\newcommand{\Ls}{L_{sg}}   
\def\g{\Gamma}

\def\s#1{{\scriptscriptstyle #1}}

\newcommand{\fatg}{{\rm{I}}\!\Gamma}

\def\tlambda{\mkern 2mu\widetilde{\mkern-4mu \lambda \mkern-2mu}\mkern 1.2mu}

\def\sd{A}  

\newcommand{\gammanew}{G} 

\begin{document}

\title{Planar degeneracy of the three-gluon vertex}

\author{A.~C. Aguilar}
\affiliation{\mbox{University of Campinas - UNICAMP, Institute of Physics ``Gleb Wataghin'',} \\
13083-859 Campinas, S\~{a}o Paulo, Brazil}

\author{M.~N. Ferreira}
\affiliation{\mbox{Department of Theoretical Physics and IFIC, 
University of Valencia and CSIC},
E-46100, Valencia, Spain}

\author{J. Papavassiliou}
\affiliation{\mbox{Department of Theoretical Physics and IFIC, 
University of Valencia and CSIC},
E-46100, Valencia, Spain}

\author{L.~R. Santos}
\affiliation{\mbox{University of Campinas - UNICAMP, Institute of Physics ``Gleb Wataghin'',} \\
13083-859 Campinas, S\~{a}o Paulo, Brazil}

\begin{abstract}

We present a detailed exploration of certain outstanding features 
of the transversely-projected three-gluon vertex, using the 
corresponding Schwinger-Dyson equation in conjunction with 
key results obtained from quenched lattice simulations. 
The main goal of this study is the 
scrutiny of the approximate property denominated ``planar degeneracy'', unveiled when the Bose symmetry of the vertex is 
properly exploited. The planar degeneracy leads 
to a particularly simple parametrization of the 
vertex, reducing its kinematic dependence to 
essentially a single variable. 
Our analysis, carried out in the 
absence of dynamical quarks, 
reveals that the planar degeneracy 
is particularly accurate for 
the description of the form factor associated with the 
classical tensor, for 
a wide array of 
arbitrary kinematic 
configurations. Instead, the remaining three form factors 
display considerable violations of this property. 
In addition, and in close connection with 
the previous point, we demonstrate  
the numerical dominance of the classical form factor 
over all others, except in the vicinity of the soft-gluon 
kinematics.
The final upshot of these 
considerations is the emergence 
of a very compact description for  the 
three-gluon vertex in  general kinematics, which 
may simplify significantly nonperturbative 
applications involving this vertex.

\end{abstract}


\maketitle

\section{Introduction}
\label{sec:intro}

The vertex that describes 
the interaction of 
three gluons, known as 
the ``three-gluon vertex'',   
plays a pivotal role in the 
dynamics of Yang-Mills theories 
in general, and of Quantum Chromodynamics (QCD) in particular~\mbox{\cite{Marciano:1977su,Ball:1980ax,Davydychev:1996pb,Gracey:2011vw,Gracey:2014mpa}}. 
The last decades have witnessed considerable progress in our
understanding of the nonperturbative structure of 
the three-gluon vertex in the Landau gauge, thanks to the 
coordinated efforts 
of lattice simulations~\cite{Parrinello:1994wd,Alles:1996ka,Parrinello:1997wm,Boucaud:1998bq,Cucchieri:2006tf,Maas:2007uv,Cucchieri:2008qm,Athenodorou:2016oyh,Duarte:2016ieu,Boucaud:2017obn,Sternbeck:2017ntv, Vujinovic:2018nqc,Aguilar:2019uob,Aguilar:2021lke,Pinto-Gomez:2022brg,Catumba:2021hng,Catumba:2021yly} and 
continuous methods,  
such as Schwinger-Dyson equations (SDEs)~\cite{Alkofer:2000wg,Alkofer:2004it,Fischer:2006ub,Huber:2012zj,Aguilar:2013vaa,Blum:2014gna,Eichmann:2014xya,Williams:2015cvx,Blum:2015lsa,Huber:2018ned,Aguilar:2019jsj,Aguilar:2019kxz,Papavassiliou:2022umz,Ferreira:2023fva} and 
functional renormalization group~\cite{Mitter:2014wpa,Cyrol:2016tym,Corell:2018yil}.
Particularly noteworthy features of this vertex 
include the suppression of its strength in the low-energy domain~\cite{Huber:2012zj,Pelaez:2013cpa,Aguilar:2013vaa,Blum:2014gna,Eichmann:2014xya,Mitter:2014wpa,Williams:2015cvx,Blum:2015lsa,Cyrol:2016tym,Corell:2018yil,Huber:2018ned,Aguilar:2019jsj,Aguilar:2019kxz,Souza:2019ylx,Barrios:2022hzr,Papavassiliou:2022umz,Ferreira:2023fva}, 
its logarithmic divergence at the origin~\cite{Aguilar:2013vaa,Aguilar:2019jsj,Aguilar:2019kxz,Papavassiliou:2022umz}, 
and the displacement of its Ward identity~\cite{Aguilar:2016vin,Aguilar:2021uwa,Papavassiliou:2022wrb,Aguilar:2022thg,Ferreira:2023fva}, induced by the 
action of the Schwinger mechanism~\cite{Schwinger:1962tn,Schwinger:1962tp,Jackiw:1973tr,Jackiw:1973ha,Eichten:1974et,Smit:1974je,Cornwall:1979hz,Cornwall:1981zr,Aguilar:2008xm,Aguilar:2011xe,Ibanez:2012zk,Eichmann:2021zuv}.

Recently, a rather striking property of 
the \emph{transversely-projected} three-gluon vertex, $\overline{\fatg}^{\,\alpha\mu\nu}(q,r,p)$,  
denominated  \emph{planar degeneracy}, has received particular attention~\cite{Pinto-Gomez:2022brg,Aguilar:2022thg,Ferreira:2023fva}. 
This property  
was first discovered in the SDE analysis of~\cite{Eichmann:2014xya}, and has been 
firmly established in a recent lattice simulation that 
explored a broad array of kinematic configurations~\cite{Pinto-Gomez:2022brg}. 
The main observation may be summarized by stating that 
when $\overline{\fatg}^{\,\alpha\mu\nu}(q,r,p)$ is spanned  
in a special tensorial basis, 
the associated form factors depend 
almost exclusively on a single kinematic variable, $s^2 = 
\frac{1}{2}(q^2+r^2+p^2)$, 
which defines a plane
in the coordinate system $(q^2, r^2, p^2)$.
Thus, 
all configurations with a common $s^2$ are nearly ``degenerate'', in the sense that 
they share,  to a high degree of accuracy, the same form factors. 
In the recent quenched lattice study of~\cite{Pinto-Gomez:2022brg}, the validity 
of this property was established for the so-called \emph{bisectoral} kinematics, 
$p^2 = r^2 \neq q^2$; its generalization 
to arbitrary configurations was conjectured on the 
grounds of an inspection including  numerous  
random configurations.
As we will see in detail in what follows, 
the systematic SDE-based exploration carried out here 
demonstrates clearly that this special 
feature persists indeed for general kinematics, 
\ie configurations with arbitrary  
$q^2$, $r^2$, and $p^2$.

The analysis presented in~\cite{Pinto-Gomez:2022brg} reached an additional 
important conclusion regarding the 
relative size of the form factors 
comprising $\overline{\fatg}^{\,\alpha\mu\nu}(q,r,p)$. Specifically, 
the form factor associated with the 
tree-level (``classical'') tensor, 
$\overline{\Gamma}_{0}^{\alpha\mu\nu}(q,r,p)$,  
dominates numerically over all others.
As a result, the particularly compact structure 
\be
\overline{\fatg}^{\alpha\mu\nu}(q,r,p) \approx  \overline{\Gamma}_{0}^{\alpha\mu\nu}(q,r,p) L_{sg}(s^{2}) \,,
\label{meq}
\ee
first used in \cite{Williams:2015cvx}, 
emerges as an excellent approximation for general kinematics.  
The function 
$L_{sg}(s^{2})$ denotes the form factor associated with the 
soft-gluon limit of the three-gluon vertex 
($q=0$, $r=-p)$, and 
is rather accurately known   
from various lattice simulations~\mbox{\cite{Athenodorou:2016oyh,Duarte:2016ieu,Boucaud:2017obn,Pinto-Gomez:2022brg,Aguilar:2019uob,Aguilar:2021lke,Aguilar:2021okw}}.

In the present work we employ the SDE that 
governs the dynamics of the three-gluon 
vertex, supplemented with inputs from quenched lattice simulations, in order to scrutinize some of 
the prominent features 
that arise after these recent developments.  
The main results of this 
exploration may be summarized as follows.

{(\it i}) 
A new basis for the expansion of $\overline{\fatg}^{\alpha\mu\nu}(q,r,p)$ is constructed, 
 which, even though it differs  
 only slightly from that of~\cite{Pinto-Gomez:2022brg},  
 it improves considerably 
 the 
exactness of the  
 planar degeneracy at the level of the individual form factors. To appreciate this point, 
 note that, in the soft-gluon limit, \1eq{meq} becomes exact, since only one transverse tensor can be constructed in this configuration. The main advantage of the new basis is that only its classical tensor is 
 nonvanishing in the soft-gluon limit, leading to the equality between the associated form factor and 
 $\Ls(r^2)$.
 As a consequence, in this new basis  
 the planar degeneracy is more accurately fulfilled for small $q$,  
 since it is exact, by construction, at $q = 0$. In contrast, 
 in the basis of~\cite{Pinto-Gomez:2022brg}, 
 $\Ls(r^2)$ emerges as  a linear combination of 
 two form factors, both of which 
  deviate markedly from the planar degeneracy at small $q$.
  
({\it ii}) 
One of the prime 
objectives  of our SDE analysis is to establish the 
extent of validity of \1eq{meq}. 
To that end, the  
soft-gluon limit of the SDE is determined, and 
all fully-dressed vertices appearing in the resulting expressions are replaced by 
the Ansatz of \1eq{meq}. 
This procedure provides a dynamical equation for 
the function $L_{sg}(r^{2})$, which is solved iteratively, using 
lattice inputs for most of the remaining ingredients, 
 such as gluon and ghost 
propagators. The resulting $L_{sg}(r^{2})$ is in very good agreement 
with the lattice data of~\cite{Aguilar:2019uob,Aguilar:2021lke,Aguilar:2021okw}.
This fine coincidence indicates that the combined error originating 
from the truncation of the SDE and potential inaccuracies in the form of \1eq{meq} is rather negligible. 


  ({\it iii}) 
The simplifications induced  by \1eq{meq} are exploited 
at the level of the vertex SDE, in order to determine  
all the form factors of $\overline{\fatg}^{\alpha\mu\nu}(q,r,p)$ in  general kinematics. 
This is accomplished by simply substituting all 
three-gluon vertices appearing 
in the SDE the r.h.s. of \1eq{meq} and carrying out the corresponding integration 
(\ie no iterative procedure is employed).
Our results 
demonstrates that the classical form factor dominates over all others, including the tensor structure not evaluated in the lattice study of~\cite{Pinto-Gomez:2022brg}, for nearly 
all kinematic regions. The only exception is the soft-gluon limit, where the non-classical  form factors grow in magnitude; this is due to a would-be collinear divergence, 
which, even though tamed by the emergence of a dynamical gluon mass~\cite{Cornwall:1981zr,Halzen:1992vd,Aguilar:2002tc,Aguilar:2006gr,Aguilar:2008xm, Luna:2005nz,Binosi:2009qm,Dudal:2008sp,Oliveira:2010xc, Cucchieri:2011ig,Serreau:2012cg,Binosi:2014aea, Kondo:2014sta,Aguilar:2015bud,Gao:2017uox,Roberts:2020hiw,Horak:2022aqx,Ding:2022ows}, leads to a considerable enhancement. 
Nevertheless, since the tensors associated with these form factors vanish in this  limit, \1eq{meq} is unaffected by this observation.
In addition,  by charting regions of momenta away from the bisectoral limit, we find that 
the planar degeneracy 
persists at a notable degree of accuracy 
at the level of the classical form factor, but is 
significantly violated at the level of the remaining three 
form factors.
In particular, in the case of the 
classical form factor,  
the largest deviation from the planar degeneracy 
is \mbox{$17.5\%$} at \mbox{$q^2 = r^2 = p^2 = 2$~$\rm GeV^2$}, rapidly dropping 
below \mbox{$10\%$} away from this point. Finally, we observe that the effect of these deviations is that \1eq{meq} generally tends to \emph{underestimate} the true value of the classical form factor.

({\it iv}) 
An interesting by-product of our analysis is related 
with the origin of the infrared 
suppression displayed by the main form factors of the three-gluon vertex, 
which acquire their
tree-level value (unity) at \mbox{$4.3$~GeV} (renormalization point), but reduce their size by half at around 
\mbox{$1$~GeV}~\mbox{\cite{Aguilar:2013vaa,Athenodorou:2016oyh,Boucaud:2017obn,Blum:2015lsa,Corell:2018yil,Huber:2018ned, Aguilar:2019jsj}}. 
  The detailed evaluation of the various diagrams comprising the  
  SDE of the three-gluon vertex, employing \1eq{meq} as input,  
  leads to a  reassessment of the origin of this phenomenon.  
Specifically, the cause of the suppression has been originally 
attributed to the infrared divergence of the ``unprotected'' logarithm 
stemming from the ghost loop diagram~\cite{Aguilar:2013vaa,Papavassiliou:2022umz}. 
However, the present analysis reveals that the contribution from the 
ghost loop becomes discernible only below \mbox{$0.5$~GeV}, converting the  
so-called ``swordfish'' diagrams (whose logarithms are ``protected'' 
by the gluon mass) to the main source of the suppression. Consequently, 
the true infrared divergence becomes apparent considerably deeper in the infrared than originally thought, in a region of momenta not accessible to current lattice simulations~\cite{Aguilar:2021lke,Boucaud:2002fx,Boucaud:2003xi}.

The article is organized as follows. 
In Sec.~\ref{sec:background} we discuss general features of $\overline{\fatg}^{\alpha\mu\nu}(q,r,p)$, placing special emphasis on the properties of the tensor basis used to decompose the vertex. Then, in Sec.~\ref{sec:SDE3g} we present the SDE governing the evolution of $\overline{\fatg}^{\alpha\mu\nu}(q,r,p)$,
derived from the 
three-particle irreducible
(3PI) three-loop effective action, 
and discuss its renormalization. 
 In Sec.~\ref{sec:softgluon} we solve this SDE in the soft-gluon limit, and compare the result with the $\Ls(r^2)$ obtained from the lattice. Next, in Sec.~\ref{sec:suppression} we discuss the 
 true  origin of the infrared suppression of 
 $\Ls(r^2)$, as unraveled through the  diagram-by-diagram evaluation of the SDE.  
 In Sec.~\ref{sec:genkin} we carry out the SDE analysis for completely general kinematics. In particular,  
 we determine  all form factors of $\overline{\fatg}^{\alpha\mu\nu}(q,r,p)$,  
analyze in detail their size hierarchy,  and the degree of accuracy of the planar degeneracy displayed by the 
classical form factor.  Finally, our conclusions are summarized in Sec.~\ref{conc}, while certain technical details are relegated to an Appendix.
%

\section{Planar degeneracy and form factor hierarchies}
\label{sec:background}

\begin{figure}[t!]
\includegraphics[width=0.4\linewidth]{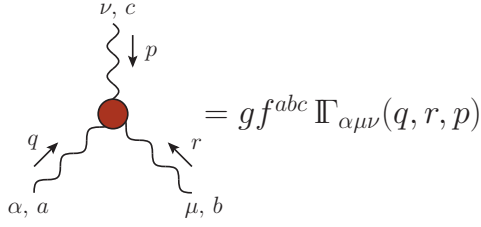}	
\caption{ Diagrammatic representation of the three-gluon vertex.}
\label{3g_def}
\end{figure}

The full three-gluon vertex is represented diagrammatically in \fig{3g_def} and denoted by 
\be
\fatg^{abc}_{\alpha\mu\nu}(q,r,p) = g f^{abc} \fatg_{\alpha\mu\nu}(q,r,p)\,,
\label{defvert}
\ee
where $g$ is the gauge coupling, and $f^{abc}$ are the SU(3) structure constants. The vertex $\fatg^{abc}_{\alpha\mu\nu}(q,r,p)$ possesses full Bose symmetry, remaining invariant under the exchange of any two sets of indices, 
\eg $(a, \alpha, q) \leftrightarrow (c, \nu, p)$. 
As has been explained in the related literature~\mbox{\cite{Ball:1980ax,Eichmann:2014xya,Aguilar:2019jsj,Pinto-Gomez:2022brg}}, this 
particular symmetry imposes numerous constraints on the 
structure of the form factors comprising this vertex~(see, \eg Eqs.~(3.7)-(3.10) in~\cite{Aguilar:2019jsj}).

At tree-level,  $\fatg^{\alpha\mu\nu}(q,r,p)$ acquires the standard expression  
\be
\Gamma_{\!0}^{\alpha\mu\nu}(q,r,p)  = (q-r)^{\nu}g^{\alpha\mu} + (r-p)^{\alpha}g^{\mu\nu} + (p-q)^{\mu}g^{\alpha\nu}\,. \label{3g_tree}
\ee
The perturbative aspects of $\fatg^{\alpha\mu\nu}(q,r,p)$ have been 
explored in various articles, see, \eg~\cite{Celmaster:1979km,Ball:1980ax,Davydychev:1996pb,Davydychev:1997vh,Gracey:2019mix,Gracey:2011vw,Gracey:2014mpa}.


The focal point of the present work is the transversely-projected vertex, $\overline{\fatg}^{\,\alpha\mu\nu}(q,r,p)$, defined as 
\be
\overline{\fatg}^{\,\alpha\mu\nu}(q,r,p) := P_{\alpha'}^{\alpha}(q) P_{\mu'}^{\mu}(r) P_{\nu'}^{\nu}(p) \fatg^{\alpha'\!\mu'\!\nu'}(q,r,p) \,, \qquad P_{\mu\nu}(q) := g^{\mu\nu} - \frac{q^\mu q^\nu}{q^2} \,.
\label{eq:Gammabar}
\ee
Evidently, \mbox{$f^{abc}\,\overline{\fatg}^{\,\alpha\mu\nu}(q,r,p)$} 
displays also full Bose symmetry, which will play a key role in what follows. 
Its tree-level value, to be denoted by $\overline{\g}_{\!0}^{\,\alpha \mu \nu}(q,r,p)$, is obtained from \1eq{eq:Gammabar} through the 
substitution $\fatg^{\,\alpha\mu\nu}(q,r,p) \to \Gamma_{\!0}^{\,\alpha\mu\nu}(q,r,p)$.

In general kinematics, \mbox{$\overline{\fatg}^{\,\alpha\mu\nu}(q,r,p)$} can be decomposed in terms of four independent tensors, \ie
\be
\overline{\fatg}^{\,\alpha \mu \nu}(q,r,p) = \sum_{i=1}^4 {\gammanew}_i(q^2,r^2,p^2) \, \tau_{i}^{\alpha\mu\nu}(q,r,p)\,,
\label{new_projection2}    
\ee
where the ${\gammanew}_i(q^2,r^2,p^2)$ denote scalar form factors, which depend on three Lorentz scalars.

Remarkably, lattice~\cite{Pinto-Gomez:2022brg,Pinto-Gomez:2022qjv,Pinto-Gomez:2023lbz} and continuum studies~\cite{Eichmann:2014xya,Blum:2014gna,Huber:2016tvc} have revealed that with a suitable choice of basis tensors, to be 
denoted by $\tau_i^{\alpha \mu \nu}$, the structure of the form factors $G_i$ is dramatically simplified. Specifically, the basis is required to satisfy the following properties:
\begin{enumerate}[label=({\itshape\roman*})]
\item All $\tau_i^{\alpha \mu \nu}(q,r,p)$ are antisymmetric under the exchange of any pair of external legs of the vertex, \eg
\be 
\tau_i^{\alpha \mu \nu}(q,r,p) = - \tau_i^{\mu \alpha \nu}(r,q,p) \,.
\ee
Consequently, the Bose symmetry of the vertex manifests itself 
in a particularly transparent way 
at the level of the individual form factors. Specifically, all $\gammanew_i(q^2,r^2,p^2)$ are symmetric under the exchange of any pair of momenta.

\item The tensor  $\tau_1^{\alpha \mu \nu}(q,r,p)$ is 
chosen to be the classical Lorentz structure,
\be 
\tau_1^{\alpha\mu\nu}(q,r,p) = \overline{\g}_{\!0}^{\,\alpha \mu \nu}(q,r,p) \,;
\ee
thus, at tree-level, the form factors reduce to  \mbox{${\gammanew}_{\!1}^{0}= 1$} and \mbox{${\gammanew}_{\!j}^{0}=0$}, for \mbox{$j=2,3,4$}. 

\item Each tensor $\tau_i^{\alpha\mu\nu}(q,r,p)$ has 
mass dimension one, exactly as the 
vertex itself. Hence, the form factors $\gammanew_i(q^2,r^2,p^2)$ are all dimensionless and can be directly compared to one another.

\end{enumerate}

Given such a basis, it is clear from 
property $(i)$ above that the form factors $G_i(q^2,r^2,p^2)$ can only depend on three Bose symmetric combinations of the momenta. Importantly, lattice results~\cite{Pinto-Gomez:2022brg,Pinto-Gomez:2022qjv,Pinto-Gomez:2023lbz} have shown that for a large range of kinematic configurations, these form factors can be accurately approximated by functions of a single Bose symmetric variable,
\be 
G_1(q^2,r^2,p^2) \approx G_1(s^2) \,, 
\label{planar}
\ee
where\footnote{In~\cite{Eichmann:2014xya}, the variable $s^2$, 
and the angles $\alpha$ and $\beta$ that we use in Sec.~\ref{subsec:planardeg}, 
were obtained through permutation group methods. }
\be
s^{2} := \frac{1}{2}(q^2 + r^2 + p^2)\,. 
\label{s_variable}
\ee
\1eq{planar} defines the property called ``planar degeneracy''. 
Note that in~\cite{Pinto-Gomez:2022brg} 
the validity of \1eq{s_variable} was proposed also 
for the remaining form factors,  \ie $G_{2,3,4}$; 
however, as we will see in Sec.~\ref{subsec:planardeg}, this is a rather poor approximation.

Furthermore, $G_{2,3,4}$ are found to be subleading 
compared to $G_1$,  for most kinematic configurations. Evidently, this property holds in perturbation theory, but its 
validity in the nonperturbative regime is 
less obvious.

The upshot of the above two observations is that $\overline{\fatg}^{\,\alpha \mu \nu}(q,r,p)$ can be accurately approximated by
\be 
\overline{\fatg}^{\,\alpha \mu \nu}(q,r,p) \approx G_1(s^2) \overline{\g}_{\!0}^{\,\alpha \mu \nu}(q,r,p) \,,
\label{compact}
\ee
over a wide range of kinematic configurations.

Now, in the soft-gluon limit, $q \to 0$, there exists only one transverse tensor structure for the three-gluon vertex~\cite{Aguilar:2021okw}, namely the tree-level vertex evaluated at $q \to 0$,
\be 
\lim_{q\to 0} \overline{\g}_{\!0}^{\,\alpha \mu \nu}(q,r,p) = 2 r^{\alpha'}P^{\mu\nu}(r) \lim_{q \to 0} P^\alpha_{\alpha'}(q) \,. \label{sg_tree_level}
\ee
Hence, the transverse vertex reduces \emph{exactly} to
\be 
\overline{\fatg}^{\,\alpha \mu \nu}(0,r,-r) = \Ls(r^2) \lim_{q\to 0} \overline{\g}_{\!0}^{\,\alpha \mu \nu}(q,r,p)\,, \label{sg_tens}
\ee
where the single form factor $\Ls(r^2)$ can be determined as~\cite{Athenodorou:2016oyh,Boucaud:2017obn,Aguilar:2021lke,Aguilar:2021okw}
\begin{align}
\label{eq:Lsg}
\Ls(r^2) =  \frac{{\Gamma}_{0}^{\alpha\mu \nu}(q,r,p)
\overline{\fatg}_{\alpha\mu\nu}(q,r,p)}
{\rule[0cm]{0cm}{0.45cm}\; {{\Gamma}_{0}^{\alpha\mu\nu}(q,r,p)  \overline{\Gamma}_{0\,\alpha\mu\nu}(q,r,p)}}
\rule[0cm]{0cm}{0.5cm} \Bigg|_{\substack{\!\!q\to 0 \\ p\to -r}} \,. 
\end{align}
At tree-level, $\Ls^{0} = 1$. Note that the limit $q\to 0$ of the term $P^\alpha_{\alpha'}(q)$ in \1eq{sg_tens} is finite, but path-dependent; nevertheless, 
as explicitly shown in \cite{Aguilar:2021uwa},
the path-dependence cancels in the ratio of \1eq{eq:Lsg}, leading to a well-defined $\Ls(r^2)$.

At this point, if the range of validity of \1eq{planar} includes the soft-gluon limit, $q = 0$, then
\be 
G_1(q^2,r^2,p^2) \approx \Ls(s^2) \,. 
\label{planar_L}
\ee
Under this additional assumption, \1eq{compact} can be recast in the form given by \1eq{meq}. The advantage of this latter expression is that the soft-gluon form factor, $\Ls(s^2)$, has been extensively studied in large-volume lattice simulations, 
appropriately refined to eliminate scale-setting and continuum extrapolation artifacts~\cite{Athenodorou:2016oyh,Duarte:2016ieu,Boucaud:2017obn,Aguilar:2019uob,Aguilar:2021lke,Aguilar:2021okw}. Consequently, the shape and size of $\Ls(s^2)$ are currently rather well-known; 
therefore, \1eq{meq} serves as a compact and accurate approximation for the general kinematics $\overline{\fatg}^{\alpha\mu\nu}(q,r,p)$.

A tensor basis that satisfies all of the conditions $(i)$--$(iii)$ above is given by
\begin{align}
\begin{split}
\tau_1^{\alpha \mu \nu} =& \,
\overline{\Gamma}_{\!0}^{\,\alpha \mu\nu}
\,,\\ 
\tau_2^{\alpha \mu \nu} =& \, \frac{3}{2 s^2} \,(q-r)^{\nu'} (r-p)^{\alpha'} (p-q)^{\mu'} 
P_{\alpha'}^\alpha(q)  P_{\mu'}^{\mu}(r)  P_{\nu'}^\nu(p)\,,\\
\tau_3^{\alpha \mu \nu} =& - \tau_2^{\alpha \mu \nu}/4 - 3 t_4^{\alpha\mu\nu}/s^2 \,,\\
\tau_4^{\alpha \mu \nu}  =& \,\left( \frac{3}{2 s^2}\right)^{\!\!2}
\left[t_1^{\alpha\mu\nu} + t_2^{\alpha\mu\nu} + t_3^{\alpha\mu\nu}\right]\,,
\end{split}
\label{lambdaBasis}
\end{align}
where we suppress the functional dependence, $(q,r,p)$, of all tensors for compactness, and the $t_i^{\alpha\mu\nu}$ are given by~\cite{Aguilar:2019jsj}
\begin{align}
t_1^{\alpha\mu\nu} =& [(q\cdot r)g^{\alpha\mu} - q^{\mu}r^\alpha][(r\cdot p)q^\nu - (q\cdot p)r^\nu]\,,
\nonumber\\
t_2^{\alpha\mu\nu} =& [(r\cdot p)g^{\mu\nu} - r^{\nu}p^\mu][(p\cdot q)r^\alpha - (r\cdot q)p^\alpha]\,,
\nonumber\\
t_3^{\alpha\mu\nu} =& [(p\cdot q)g^{\nu\alpha} - p^{\alpha}q^\nu][(q\cdot r)p^\mu - (p\cdot r)q^\mu]\,,
\nonumber\\
t_4^{\alpha\mu\nu} =& g^{\mu\nu}[ (r\cdot q)p^\alpha - (p\cdot q)r^\alpha ] + g^{\nu\alpha}[ (p\cdot r)q^\mu - (q\cdot r)p^\mu ] + g^{\alpha\mu}[ (q\cdot p)r^\nu - (r\cdot p)q^\nu ]
\nonumber\\
& + r^\alpha p^\mu q^\nu - p^\alpha q^\mu r^\nu \,.
\label{ti}
\end{align}

The basis given in \1eq{lambdaBasis} 
is a minimal modification of the basis
employed in~\cite{Pinto-Gomez:2022brg,Pinto-Gomez:2023lbz}, 
where $\overline{\fatg}^{\,\alpha \mu \nu}(q,r,p)$ 
was decomposed as 
\begin{align}
\overline{\fatg}^{\,\alpha \mu \nu}(q,r,p) = \sum_{i=1}^4 \widetilde{\Gamma}_i(q^2,r^2,p^2) \,
\tlambda_i^{\alpha\mu\nu}(q,r,p) \,;
\label{eq:expl}    
\end{align}
the tensors $\tlambda_i^{\alpha\mu\nu}(q,r,p)$ are related to the $\tau^{\alpha\mu\nu}_{j}(q,r,p)$ through the simple relations 
\begin{align}
\label{eq:changeBasis}
\tau^{\alpha\mu\nu}_{j}(q,r,p) &= \tlambda_{j}^{\alpha \mu \nu}(q,r,p)\,,   \qquad j=1,2,4\,
, \nonumber \\
\tau^{\alpha\mu\nu}_{3}(q,r,p) &= \tlambda_{3}^{\alpha \mu \nu}(q,r,p) - \tfrac{3}{2}\,\tlambda_{1}^{\alpha \mu \nu}(q,r,p)\,.
\end{align} 
Hence, the form factors  $\gammanew_i(q^2,r^2,p^2)$ and $\widetilde{\Gamma}_i(q^2,r^2,p^2)$ of both bases are connected through the transformations
 \begin{align}
\label{eq:newformfactors}   
 \gammanew_1(q^2,r^2,p^2)  &= \widetilde{\Gamma}_{1}(q^2,r^2,p^2)  + \tfrac{3}{2}\widetilde{\Gamma}_{3}(q^2,r^2,p^2) \,, \nonumber \\
\gammanew_j(q^2,r^2,p^2)  &= \widetilde{\Gamma}_{j}(q^2,r^2,p^2)\,,  \qquad j=2,3,4\,.
\end{align}

The motivation for using \1eq{lambdaBasis}, instead of \1eq{eq:changeBasis}, is that, in the soft-gluon limit, the two tensors $\tlambda_{1,3}^{\alpha\mu\nu}(q,r,p)$ survive and become linearly dependent, \ie 
\be 
(2/3)\lim_{q\to 0}\tlambda_3^{\alpha\mu\nu}(q,r,p) = \lim_{q\to 0}\tlambda_1^{\alpha\mu\nu}(q,r,p) = \lim_{q\to 0}\overline{\Gamma}_{\!0}^{\,\alpha \mu\nu}(q,r,p) \,,
\ee
such that
\begin{align}
\label{eq:old_basis_sg_limit}
\tlambda_1^{\alpha\mu\nu}(0,r,-r) &=  (2/3) \tlambda_3^{\alpha\mu\nu}(0,r,-r) = 2 r^{\alpha'}P^{\mu\nu}(r) \lim_{q \to 0} P^\alpha_{\alpha'}(q)\,, \nonumber \\
\tlambda_j^{\alpha\mu\nu}(0,r,-r) &= 0\,,   \qquad j=2,4\,.
\end{align} 
Hence, in the basis $\{\tlambda_i\}$,  
$\Ls(r^2)$ is not obtained as the limit 
of $\widetilde{\Gamma}_{1}(q^2,r^2,p^2)$ as $q \to 0$, 
but rather as a linear combination of $\widetilde{\Gamma}_{1}(0,r^2,r^2)$ and $\widetilde{\Gamma}_{3}(0,r^2,r^2)$. 
However,  as we will show in Sec.~\ref{sec:genkin}, $G_3(q^2,r^2,p^2)$, and hence $\widetilde{\Gamma}_{3}(q^2,r^2,p^2)$, is sizable at $q = 0$; such that the equivalent of \1eq{planar_L} with $G_1$ substituted by $\widetilde{\Gamma}_{1}$ is a poor approximation in the soft-gluon limit.

In contrast, in the basis of \1eq{lambdaBasis}, only $\tau_1^{\alpha\mu\nu}$ is nonvanishing at $q = 0$, \ie
\begin{align}
\label{eq:basis_sg_limit}
\tau_1^{\alpha\mu\nu}(0,r,-r) &= 2 r^{\alpha'}P^{\mu\nu}(r) \lim_{q \to 0} P^\alpha_{\alpha'}(q)\,, \nonumber \\
\tau_j^{\alpha\mu\nu}(0,r,-r) &= 0\,,   \qquad j=2,3,4\,.
\end{align} 
Consequently, comparing \2eqs{sg_tens}{eq:basis_sg_limit}, we obtain simply
\be
\gammanew_1(0,r^2,r^2)= \Ls(r^2) \,, \label{eq:gammaLsg}
\ee
\ie \1eq{planar_L} is promoted to an exact relation in the soft-gluon limit.
As a result, we find that the property of the 
planar degeneracy, as captured by \1eq{planar_L},  is more accurately realized with the basis of \1eq{lambdaBasis} than with that of 
\1eq{eq:expl}.  

Armed with these observations, in the rest of this article we will 
focus on three key 
issues, namely 
the derivation of $\Ls(r^2)$ from the vertex SDE, 
the dominance of $\gammanew_1(q^2,r^2,p^2)$
over the other three form factors, 
and the quantification 
of the planar degeneracy 
(or the deviations from it) 
in the case of $\gammanew_1(q^2,r^2,p^2)$, 
for general kinematics.

\section{SDE of the three-gluon vertex}
\label{sec:SDE3g}

In this section we present the 
SDE of the three-gluon vertex that we will 
employ in our analysis, 
and discuss in detail how its 
renormalization is implemented. 

\begin{figure}[ht]
 \includegraphics[width=0.95\linewidth]{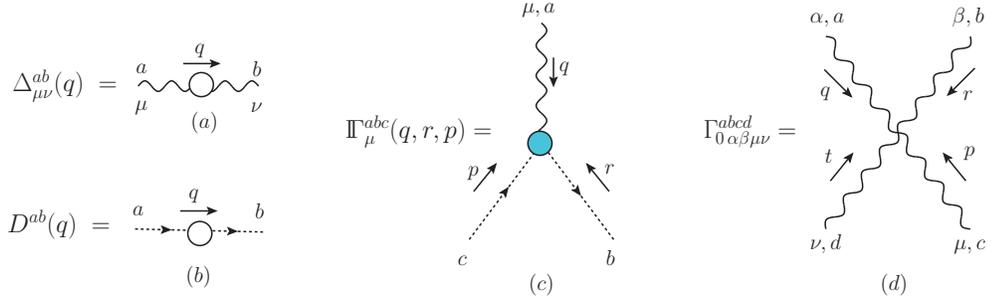}
 \caption{Diagrammatic representations of: (a) the fully dressed gluon propagator,  $\Delta^{ab}_{\mu\nu}(q)$;  (b) the complete ghost propagator, $D^{ab}(q)$; (c) the full ghost-gluon vertex, $\fatg^{abc}_{\mu}(q,r,p)$; (d) the tree-level four-gluon vertex, $\Gamma_{\!0\,\alpha \beta \mu \nu}^{abcd}$.} 
\label{fig:prop_vert}
\end{figure}

Throughout this work we adopt the version of the three-gluon SDE 
obtained within the formalism of the 3PI 
effective action~\cite{Cornwall:1974vz,Cornwall:1973ts}, at the {\it three-loop} level
~\cite{Berges:2004pu,Carrington:2010qq,York:2012ib,Carrington:2013koa,Williams:2015cvx}; for general reviews on the SDE formalism, see, \eg~\cite{Roberts:1994dr,Fischer:2006ub,Roberts:2007ji,Cloet:2013jya,Huber:2018ned,Roberts:2020hiw}.

\begin{figure}[ht]
 \includegraphics[width=0.95\linewidth]{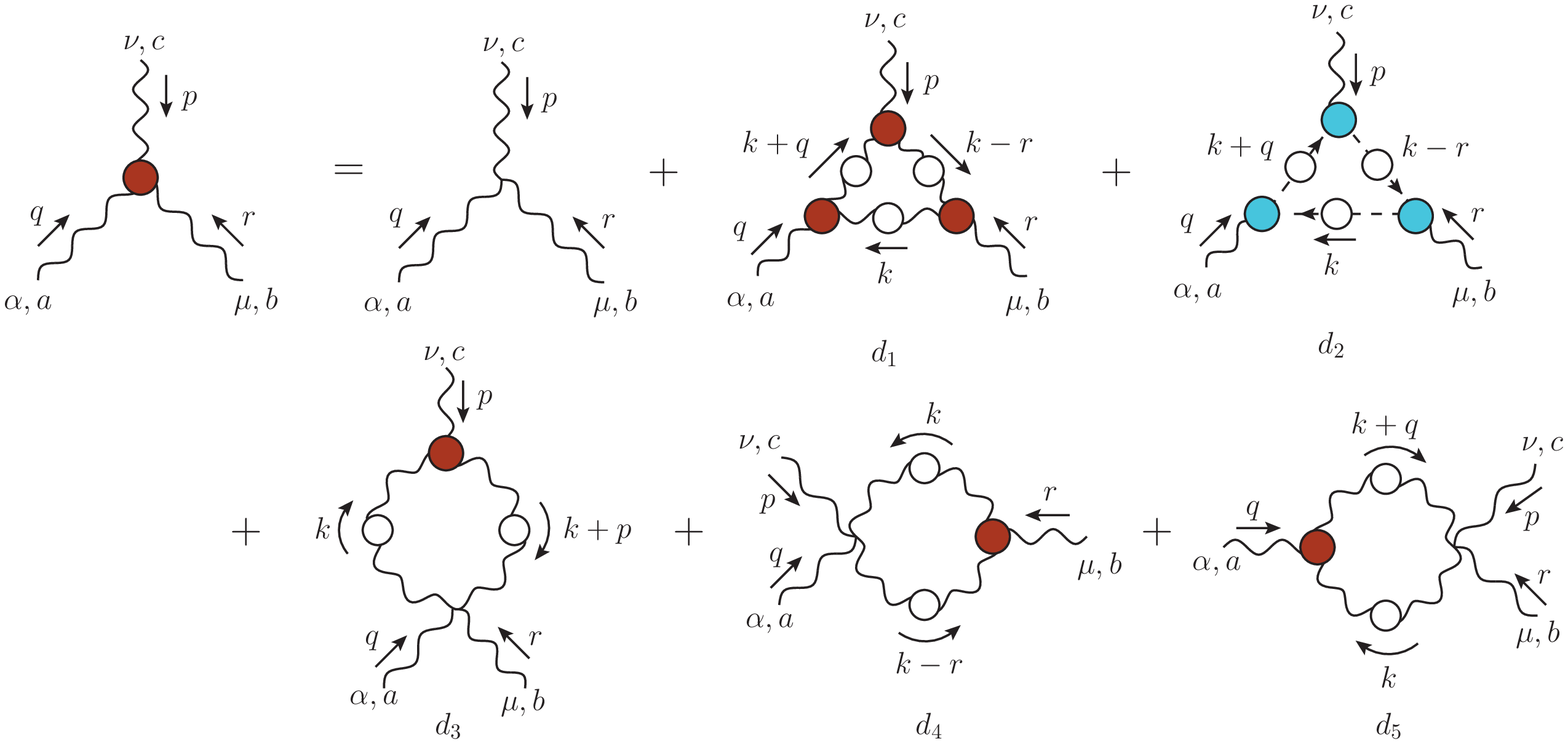}
 \caption{Diagrammatic representation of the SDE for the  full three-gluon vertex,  $\fatg^{\alpha\mu\nu}$,  derived from the 
 3PI effective action at the three-loop level. Diagrams $d_3$, $d_4$, and, $d_5$ are often referred to as ``swordfish diagrams''. The ghost loop with the arrow reversed 
 (equal in contribution to $d_2$) is not shown.}
\label{fig:SDE}
\end{figure}

As has been explained in the related literature,
the diagrammatic representations of the 
SDEs originates from 
the functional differentiation of 
this particular action (see Figs.~1-2 of \cite{Williams:2015cvx}), and  
involves fully dressed two- and three-point functions
(\ie gluon and ghost propagators, and three-gluon and  
ghost-gluon vertices). On the other 
hand, the four-gluon vertex, \mbox{$\fatg_{\alpha \beta \mu \nu}^{abcd}(q,r,p,t)= -ig^2 \fatg_{\alpha \beta \mu \nu}^{abcd}(q,r,p,t)$}, 
receives no quantum corrections, thus 
retaining its tree-level (classical) 
form, given by\footnote{The dressed version of this latter vertex 
appears when the 4PI effective action
is employed (see, \eg \cite{Carrington:2010qq}).}
\begin{align}
&\Gamma^{abcd}_{\!0\,\alpha\beta\mu\nu} =  f^{adx}f^{cbx}\left(g_{\alpha\mu}g_{\beta\nu}-g_{\alpha\beta}g_{\mu\nu}\right) 
		+f^{abx}f^{dcx}\left(g_{\alpha\nu}g_{\beta\mu}-g_{\alpha\mu}g_{\beta\nu}\right) \nonumber \\ 
		& \quad \quad +f^{acx}f^{dbx}\left(g_{\alpha\nu}g_{\beta\mu}-g_{\alpha\beta}g_{\mu\nu}\right) \,. 
\end{align}
The diagrammatic representations of all the quantities mentioned above 
are given in Figs.~\ref{3g_def} and~\ref{fig:prop_vert}. In terms of these
components, the resulting SDE for the three-gluon vertex acquires the form 
shown in \fig{fig:SDE}. Our analysis is restricted to the case of pure Yang-Mills, 
with no active quark flavours, hence the 
absence of quark loops in \fig{fig:SDE}. 

To determine the transversely-projected three-gluon vertex, ${\overline \fatg}^{\alpha\mu\nu}$, defined in \1eq{eq:Gammabar}, the first step is to contract each external leg of the SDE of \fig{fig:SDE} with a corresponding transverse projector. Then, since we specialize to the \emph{Landau gauge}, the gluon propagator $\Delta^{ab}_{\mu\nu}(q)=-i\delta^{ab}\Delta_{\mu\nu}(q)$ assumes the completely transverse form, 
\begin{align}
\Delta_{\mu\nu}(q) = \Delta(q^2)P_{\mu\nu}(q)\,.
\end{align}
Therefore, the legs of the three-gluon vertices attached to internal lines are automatically projected transversely as well.
Consequently, in the Landau gauge, 
the SDE of \fig{fig:SDE} furnishes a self-consistent equation for ${\overline \fatg}^{\alpha\mu\nu}$, without reference to longitudinal tensor structures~\cite{Eichmann:2014xya,Huber:2018ned}.

In terms of the diagrams shown in \fig{fig:SDE}, 
the transversely-projected 
three-gluon vertex is given by 
\be
\label{3gsde}
\overline{\fatg}^{\,\alpha\mu\nu}(q,r,p) = \overline{\Gamma}_{\!0}^{\,\alpha \mu \nu}(q,r,p) \,
+ \, \sum_{i=1}^5 \bar{d}_i^{\,\alpha\mu\nu}(q,r,p) \,, 
\ee
where 
\be
\bar{d}_i^{\,\alpha\mu\nu}(q,r,p) := P_{\alpha'}^{\alpha}(q) P_{\mu'}^{\mu}(r) P_{\nu'}^{\nu}(p)  \,{d}_i^{\,\alpha'\mu'\nu'}
(q,r,p) \,.
\label{dbar}
\ee

After carrying out the color algebra, the individual 
contributions of the 
graphs ${d}_i$ in Minkowski space are given by 
%
\begin{align}
\label{dis}
\bar{d}_1^{\,\alpha\mu\nu}(q,r,p) &= \lambda \!\! \int_k  \overline{\fatg}^{\alpha\sigma\beta}(q,-k_1, k) \overline{\fatg}^{\mu\beta\tau}(r,-k, k_{2}) \overline{\fatg}^{\nu\tau\sigma}(p,-k_2,k_1) \Delta(k_1^2) \Delta(k^2) \Delta(k_2^2)\,, \nonumber \\
\bar{d}_2^{\,\alpha\mu\nu}(q,r,p) &= - 2\lambda \!\! \int_k \overline{\fatg}^{\alpha}(q,-k_1,k)\overline{\fatg}^{\mu}(r,-k,k_2)\overline{\fatg}^{\nu}(p,-k_2,k_1) D(k_1^2)D(k^2)D(k_2^2) \,, \nonumber  \\
\bar{d}_3^{\,\alpha\mu\nu}(q,r,p)&=  \frac{3}{2} \lambda  [P^{\alpha}_{\sigma}(q)P^{\mu}_{\beta}(r) -P^{\alpha}_{\beta}(q) P^{\mu}_{\sigma}(r)]\!\!\int_k 
\overline{\fatg}^{\nu\sigma\beta}(p,-k_3,k) 
\Delta(k_3^2) \Delta(k^2) \,, \nonumber
\\
\bar{d}_4^{\,\alpha\mu\nu}(q,r,p) &= -\bar{d_3}^{\,\alpha\nu\mu}(q,p,r) \,, 
\nonumber \\
\bar{d}_5^{\,\alpha\mu\nu}(q,r,p) &= -\bar{d_3}^{\,\nu\mu\alpha}(p,r,q) \,,
\end{align}
where we define \mbox{$k_1=k+q$}, \mbox{$k_2=k-r$}, and \mbox{$k_3=k+p$}, together with $ \lambda := ig^{2}C_{\rm A}/2$, with $C_\mathrm{A}$ denoting the eigenvalue of the Casimir operator in the adjoint representation 
[$C_\mathrm{A} = N$ for SU($N$)]. In addition, we denote by \mbox{ $D^{ab}(q)= i\delta^{ab}D(q^2)$} the ghost propagator, whose dressing function,  $F(q^2)$, is given by  
\be
\label{eq:ghost_dressing}
D(q^2) =\frac{F(q^2)}{q^2}\,.
\ee
Furthermore, $\fatg^{abc}_\mu(q,r,p) =-g f^{abc} \fatg_\mu(q,r,p)$ stands for the ghost-gluon vertex, whose Lorentz decomposition reads
\be 
\fatg_\mu(q,r,p) = r_\mu B_1(r,p,q) + q_\mu B_2(r,p,q) \,;
\ee
at tree-level, $B_1^0 = 1$ and $B_2^0 = 0$. Note that only $B_1$ contributes to the transversely-projected form, $\overline{\fatg}^{\,\mu}(q,r,p)$, defined as
\be
\overline{\fatg}^{\,\mu}(q,r,p) := P_{\mu'}^{\mu}(q) \fatg^{\mu'}(q,r,p) = r^{\mu'}P_{\mu'}^{\mu}(q)B_1(r,p,q) \,.
\ee
Finally, we denote by 
\be\label{eq:int_measure}
\int_{k} := \frac{1}{(2\pi)^4} \int \!\!{\rm d}^4 k 
\ee
the integration over virtual momenta, 
where the use of a symmetry-preserving regularization scheme is implicitly assumed.

All quantities appearing in \1eq{dis} are bare (unrenormalized). 
The transition to renormalized quantities is implemented 
by means of the standard relations 
\begin{align} 
\Delta_{\s R}(q^2)&= Z^{-1}_{A} \Delta(q^2)\,, &\quad\quad\quad \fatg^{\mu}_{\!\!\s R}(q,p,r) &= Z_1 \fatg^{\mu}(q,p,r)\,,\nonumber\\  
F_{\!\s R}(q^2)&= Z^{-1}_{c} F(q^2)\,,&\quad\quad\quad \fatg^{\alpha\mu\nu}_{\!\!\s R}(q,r,p) &=  Z_3 \fatg^{\alpha\mu\nu}(q,r,p)\,,\nonumber\\ 
g_{\s R} &= Z_g^{-1} g\,, &\quad\quad\quad  \fatg^{abcd}_{\!\!\s R\,\alpha\beta\mu\nu}(q,r,p,t) &=  Z_4 \fatg^{abcd}_{\alpha\beta\mu\nu}(q,r,p,t)\,,
\label{renconst1}
\end{align} 
where the subscript ``$R$'' denotes renormalized quantities, and  $Z_{A}$, $Z_{c}$, $Z_{1}$, $Z_{3}$, $Z_{4}$, and $Z_g$ are the corresponding (cutoff-dependent) renormalization constants.
In addition, we employ the exact relations 
\be
Z_g^{-1} = Z_1^{-1} Z_A^{1/2} Z_c\, = Z_3^{-1} Z_A^{3/2} \, = Z_4^{-1/2} Z_A \,,
\label{eq:sti_renorm}
\ee 
which are imposed by the various Slavnov-Taylor identities~\cite{Taylor:1971ff,Slavnov:1972fg}. 
Substituting the relations of \1eq{renconst1} into \1eq{dis}
and using \1eq{eq:sti_renorm}, it is straightforward to 
 derive 
the renormalized version of \1eq{3gsde}, given by 
\begin{align}
\label{renorm_3gsde}
\overline{\fatg}_{\!\!\s R}^{\,\alpha\mu\nu}(q,r,p) = Z_3\overline{\Gamma}_{\!0}^{\,\alpha \mu \nu}(q,r,p) 
\,+ \sum_{i=1}^2 \bar{d}^{\,\alpha\mu\nu}_{i,\s R}(q,r,p) 
+ Z_4
\sum_{i=3}^5 \bar{d}^{\,\alpha\mu\nu}_{i,\s R}(q,r,p) \,,
\end{align}
where the 
subscript ``R'' in $\bar{d}_{i,\s R}$ 
denotes that the expressions given in 
\1eq{dis} have been substituted by their renormalized counterparts. 
Of course, when the momentum integration over $k$ 
is carried out, 
the $\bar{d}_{i,\s R}$ diverge; their combined 
divergence will be removed subtractively, by adjusting appropriately the vertex renormalization constant $Z_3$. Instead, the multiplicative 
$Z_4$ will be approximated simply by setting $Z_4=1$; this  
is consistent with the fact that, at this level of approximation,  the four-gluon vertex receives no quantum corrections.

In order to fix the renormalization constants appearing in \1eq{renorm_3gsde}, we adopt the asymmetric MOM renormalization scheme~\cite{Aguilar:2020yni,Aguilar:2020uqw,Aguilar:2021lke,Aguilar:2020uqw}. This scheme imposes that 
\be 
\Delta^{-1}(\mu^2) = \mu^2 \,, \qquad F(\mu^2) = 1 \,, \qquad \Ls(\mu^2) = 1 \,, 
\label{ren_conds}
\ee
which means that the gluon and ghost propagators assume their tree-level values at the subtraction point  $\mu$,  while an analogous condition is imposed on the three-gluon vertex in the soft-gluon limit.

In applying the asymmetric MOM condition at the level of the SDE, we should keep in mind that the transversely-projected vertex becomes ill-defined in the soft-gluon limit. Nevertheless, the form factor $\Ls(r^2)$ is well defined, as mentioned below \1eq{eq:Lsg}, such that \1eq{ren_conds} fixes $Z_3$ uniquely. 

To see this, we simply factor out the term $P^\alpha_{\alpha^\prime}(q)$ in \1eq{renorm_3gsde}, which becomes ill-defined in the $q = 0$ limit. Then, \1eq{renorm_3gsde} becomes (with $Z_4 = 1$)
\begin{align}
\label{renorm_3gsde_PP_step1}
P_{\mu^\prime}^\mu(r)P_{\nu^\prime}^\nu(p)\fatg_{\!\!\s R}^{\,\alpha^\prime\mu^\prime\nu^\prime}(q,r,p) = P_{\mu^\prime}^\mu(r)P_{\nu^\prime}^\nu(p)\left[ Z_3\Gamma_{\!0}^{\,\alpha^\prime \mu^\prime \nu^\prime}(q,r,p) 
\,+ \sum_{i=1}^5 d^{\,\alpha^\prime\mu^\prime\nu^\prime}_{i,\s R}(q,r,p) \right] \,,
\end{align}
which is well-defined as $q \to 0$. 
Moreover, all terms in the above expression have the same tensor structure, of the form $r^{\alpha^\prime} P^{\mu\nu}(r)$, which is the only possible Lorentz structure of a generic tensor $T^{\alpha^\prime\mu\nu}(0,r,-r)$ that is transverse to $r^\mu$ and $r^\nu$. In particular
\bea 
\lim_{q\to 0} P_{\mu^\prime}^\mu(r)P_{\nu^\prime}^\nu(p)\fatg_{\!\!\s R}^{\,\alpha^\prime\mu^\prime\nu^\prime}(q,r,p) &=& 2 r^{\alpha^\prime} P^{\mu\nu}(r) \Ls(r^2) \,, \nonumber\\
\lim_{q\to 0} P_{\mu^\prime}^\mu(r)P_{\nu^\prime}^\nu(p)\Gamma_{\!0}^{\,\alpha^\prime\mu^\prime\nu^\prime}(q,r,p) &=& 2 r^{\alpha^\prime} P^{\mu\nu}(r) \,, \nonumber\\
\lim_{q\to 0} P_{\mu^\prime}^\mu(r)P_{\nu^\prime}^\nu(p) d^{\,\alpha^\prime\mu^\prime\nu^\prime}_{i,\s R}(q,r,p) &=& 2 r^{\alpha^\prime} P^{\mu\nu}(r) d_{i,\s R}^{\,\scriptscriptstyle{sg}}(r^2) \,, \label{di_sg_def}
\eea
where $d_{i,\s R}^{\,\scriptscriptstyle{sg}}(r^2)$ denotes the scalar form factors of  $P_{\mu^\prime}^\mu(r)P_{\nu^\prime}^\nu(r)d^{\,\alpha^\prime\mu^\prime\nu^\prime}_{i,\s R}(0,r,-r)$.

Then, the substitution of the above limits into \1eq{renorm_3gsde_PP_step1} yields
\begin{align}
\Ls(r^2) &=Z_{3} + \sum_{i=1}^4 d_{i,\s R}^{\,\scriptscriptstyle{sg}}(r^2)\,,
\label{contrib_Lsg}    
\end{align}
where we have used the fact that in the limit \mbox{$q\to 0$}, $d_{5}$ vanishes 
in its entirety, and, in particular, 
$d_{5,\s R}^{\scriptscriptstyle{sg}}(r^2)=0$.  In addition, notice that in this same limit, the diagrams $d_{3}$ and $d_{4}$ become equal.

By imposing the renormalization condition of \1eq{ren_conds}, it is straightforward to see that  Eq.~\eqref{contrib_Lsg} immediately determines that 
\begin{align}
   Z_{3} = 1-\sum\limits_{i=1}^4 d_{i,\s R}^{\,\scriptscriptstyle{sg}}(\mu^2)\,. 
\label{Z3_exp}    
\end{align}
Thus, substituting Eq.~\eqref{Z3_exp}  into Eq. \eqref{renorm_3gsde}, one obtains the renormalized SDE for $\overline{\fatg}^{\,\alpha\mu\nu}$, expressed as
\begin{align}
\overline{\fatg}_{\!\!\s R}^{\,\alpha\mu\nu}(q,r,p) = 
\overline{\Gamma}_{\!0}^{\,\alpha \mu \nu}(q,r,p) \left[1- \sum_{i=1}^{4} d_{i,\s R}^{\,\scriptscriptstyle{sg}}(\mu^2)\right]  + \sum_{i=1}^{5} \bar{d}_{i,\s R}^{\,\alpha\mu\nu}(q,r,p) \,.
\label{eq:finitedis}
\end{align}

It may seem at this point 
that the determination of $Z_3$ through the asymmetric scheme does not completely eliminate the divergences of the SDE, because,  while all ``swordfish'' diagrams
diverge in general kinematics, 
the $Z_3$ of \1eq{Z3_exp} is independent of $\bar{d}_5$. However, a simple 
one-loop calculation illustrates 
how \1eq{Z3_exp} cancels correctly all divergences. 
In particular, 
using dimensional regularization, with spacetime dimension \mbox{$d = 4-2\epsilon$}, we find that the divergent parts of $\bar{d}_{3,4,5}$ are given by 
\bea
\bar{d}_{3, \rm div}^{\,\alpha\mu\nu}(q,r,p) &=& \sd \, ( p_{\mu'}g_{\alpha'\nu'} - p_{\alpha'}g_{\mu'\nu'} )P^{\alpha\alpha'}(q)P^{\mu\mu'}(r)P^{\nu\nu'}(p) \,, \nonumber\\
\bar{d}_{4, \rm div}^{\,\alpha\mu\nu}(q,r,p) &=& \sd\, ( r_{\alpha'}g_{\mu'\nu'} - r_{\nu'}g_{\alpha'\mu'} )P^{\alpha\alpha'}(q)P^{\mu\mu'}(r)P^{\nu\nu'}(p) \,, \nonumber\\
\bar{d}_{5, \rm div}^{\,\alpha\mu\nu}(q,r,p) &=& \sd\, ( q_{\nu'}g_{\alpha'\mu'} - q_{\mu'}g_{\alpha'\nu'} )P^{\alpha\alpha'}(q)P^{\mu\mu'}(r)P^{\nu\nu'}(p) \,, \label{swords_1loop}
\eea
where
\be 
\sd := - \frac{1}{\epsilon}\left(\frac{15\alpha_s C_{\rm A}}{32\pi } \right) \,.
\ee
Then, we note that the sum of these divergences results in the tree-level tensor structure, for any kinematic configuration, \ie
\be 
\bar{d}_{3, \rm div}^{\,\alpha\mu\nu} + \bar{d}_{4, \rm div}^{\,\alpha\mu\nu} + \bar{d}_{5, \rm div}^{\,\alpha\mu\nu} = \sd \, \overline{\Gamma}_{\!0}^{\,\alpha \mu \nu}(q,r,p) \,. \label{swords_G1_1loop}
\ee

Next, specializing in the soft-gluon limit ($q=0$, $p = -r$), we find
\begin{align} 
\bar{d}_{3, \rm div}^{\alpha\mu\nu}(0,r,-r) =&\, \bar{d}_{4, \rm div}^{\,\alpha\mu\nu}(0,r,-r) = \sd \, r_{\alpha^\prime}P^{\mu\nu}(r) \lim_{q\to 0} P^{\alpha\alpha^\prime}(q) \,, \nonumber\\
\bar{d}_{5, \rm div}^{\,\alpha\mu\nu}(0,r,-r) =&\, 0 \,.
\end{align}
Using \1eq{di_sg_def}, it follows that the contributions of the $\bar{d}_{3,4,5}$ to the form factor $\Ls(r^2)$ are given by
\be 
d_{3, \rm div}^{\,\scriptscriptstyle{sg}} = d_{4, \rm div}^{\,\scriptscriptstyle{sg}} = \frac{\sd}{2} \,, \quad d_{5, \rm div}^{\,\scriptscriptstyle{sg}} = 0 \,, \,\,\Longrightarrow\,\, d_{3,\rm div}^{\,\scriptscriptstyle{sg}} + d_{4,\rm div}^{\,\scriptscriptstyle{sg}} + d_{5,\rm div}^{\,\scriptscriptstyle{sg}} = \sd \,.
\ee  
Finally, using the above expressions in \1eq{Z3_exp}, we see that the contribution of $d_{3, \rm div}^{\,\scriptscriptstyle{sg}}$ and $d_{4, \rm div}^{\,\scriptscriptstyle{sg}}$ to the combination $Z_3\overline{\Gamma}_{\!0}^{\,\alpha \mu \nu}(q,r,p)$ is precisely the negative of $\bar{d}_{3, \rm div} + \bar{d}_{4, \rm div} + \bar{d}_{5, \rm div}$ in general kinematics, given in \1eq{swords_G1_1loop}. Hence, the $Z_3$ defined through the asymmetric MOM scheme captures the correct divergence of all swordfish diagrams, $\bar{d}_{3,4,5}(q,r,p)$, despite the vanishing of $\bar{d}_5(0,r,-r)$, thus ensuring the finiteness of the renormalized vertex in general kinematics.

In the following sections, we perform different projections of Eq.~\eqref{eq:finitedis} 
in order to extract different kinematic limits. When no ambiguity can arise we drop the index ``R'' to avoid notation clutter.

\section{Soft-gluon configuration}
\label{sec:softgluon}

In this section, we consider the SDE determination of the soft-gluon form factor, $\Ls(r^2)$, from \2eqs{contrib_Lsg}{Z3_exp}, by explicitly employing 
the approximation given by \1eq{meq}.
Therefore, the level of agreement between the SDE outcome and available lattice data will constitute the first test of the accuracy of \1eq{meq}.

In order to appreciate how the  
approximation of \1eq{meq} is used in this context, 
note that
of all vertices  $\overline{\fatg}^{\,\alpha \mu \nu}(q,r,p)$ 
appearing in the SDE of \fig{fig:SDE}, only one supplies a 
form factor $\Ls(k^2)$ naturally, \ie as a result 
of triggering the first relation of 
\1eq{di_sg_def}: the vertex of diagram $d_1$ that
carries  $q$ as its external momentum\footnote{Remember that the diagram $d_5$ vanishes in the soft-gluon limit.}.  All other three-gluon vertices 
are evaluated in general kinematics even after setting $q = 0$, \eg $(r,-k,k-r)$. 
Therefore, strictly speaking, 
the general decomposition of \1eq{new_projection2} must be employed 
for all of them, inducing a dependence on all four 
form-factors $G_i\left(r^2,k^2,(k-r)^2\right)$. The use of \1eq{meq} enters at this point, 
by implementing  the substitution 
\be
\overline{\fatg}^{\alpha\mu\nu}(r,-k,k-r) 
\to \overline{\Gamma}_{0}^{\alpha\mu\nu}(r,-k,k-r) L_{sg}(s^{2}) \,,
\label{meq2}
\ee
with $s^2 = [r^2+k^2+(k-r)^2]/2 = k^2+r^2 - k\cdot r$, for all these vertices.
As a result, the only form factor related to the three-gluon vertex 
that appears on the r.h.s of the SDE is the $\Ls$. In particular,  
the SDE reduces to an integral equation for $\Ls$, of the general form 
\be 
\Ls = 1 + \int_k K_1  + \int_k \Ls K_2   + \int_k \Ls^3 K_3  \,, 
\label{schematic}
\ee
whose solution provides the momentum evolution of $\Ls(r^2)$.

Specifically, after conversion to Euclidean space and use of hyper-spherical coordinates,
and applying standard transformation rules (see, \eg Eq.~(5.1) of~\cite{Aguilar:2018csq}), we obtain from \2eqs{contrib_Lsg}{Z3_exp}
\begin{align}
\Ls(r^2) = 1 +  \sum\limits_{i=1}^4 d_{i,f}^{\,\scriptscriptstyle{sg}}(r^2,\mu^2)\,,  
\qquad  d_{i,f}^{\,\scriptscriptstyle{sg}}(r^2,\mu^2):= d_i^{\,\scriptscriptstyle{sg}}(r^2) - d_i^{\,\scriptscriptstyle{sg}}(\mu^2)\,,
\label{lsg_renorm}
\end{align} 
with
\begin{align}
\label{variosL}
     d_1^{\,\scriptscriptstyle{sg}}(x) =& 
    \frac{4\lambda'}{3}\!\!\int_{0}^{\infty}\!\!\!\!dy
    \!\!\int_{0}^{\pi}\!\!\!\!d\phi\,
    s_{\phi}^4 \,w \left[xy (c_{\phi}^2+8)\!+\!3(x^2\!+\!y^2)-\!6\sqrt{y x}c_{\phi}(x+y)\!\right]\!\Ls(y) \Ls^2(v)\Delta(u) \Delta^2(y) \,,\nonumber\\
     d_2^{\,\scriptscriptstyle{sg}}(x) =& 
    - \frac{\lambda'}{3}\int_{0}^{\infty}\!\!\!\!dy
    \int_{0}^{\pi}\!\!\!\!d\phi\, s_\phi^4(w/y)F^2(y)F(u)B_1^2(u,y,\chi)B_1(y,y,\pi)\,,\nonumber\\
    d_3^{\,\scriptscriptstyle{sg}}(x) =&\,   d_4^{\,\scriptscriptstyle{sg}}(x) = 
    -\frac{\lambda'}{2}\int_{0}^{\infty}\!\!\!\!dy
    \!\!\int_{0}^{\pi}\!\!\!\!d\phi \,s_{\phi }^4 \left({y}/{u}\right)
    \left[3x + 5y - 5\sqrt{x y}c_{\phi }\right]   \Ls(v) \Delta(u) \Delta(y)\,,
\end{align}
where $\lambda':= {C_{A}\alpha_s}/{4\pi^{2}}$ and $\alpha_s(\mu^2):=g^2/4\pi$ is the value of the strong charge at the renormalization point $\mu$. In the above equation, we introduced the auxiliary variables
\begin{align}
 \label{spherical_coord}
x &:= r^2, \qquad y := k^2, \qquad  u := (r-k)^2 = x + y - 2\sqrt{xy}c_\phi\,, \nonumber \\
v &:= x + y - \sqrt{xy}c_\phi\,, \qquad w:=\left({y}/{u}\right)\sqrt{{y}/{x}}\,c_{\phi}\,,
\end{align}
where $\phi$ denotes the angle between the momenta $k$ and $r$, while \mbox{$s_\phi := \sin \phi$}, \mbox{$c_\phi := \cos \phi$}. Furthermore, we parametrize $B_1(r,p,q)$, in terms of the squares of their first two arguments and the angle between them.  
In particular, 
\begin{align}
 \label{Bspherical}
 B_1(k,-k,0) \to  B_1(y,y,\pi) \,, 
 &&
 B_1(k-r,-k,r) \to   B_1(u,y,\chi)\,,
\end{align}
with the angle $\chi$ defined as \mbox{$c_{\chi}:=\left(\sqrt{{x}/{u}}\,c_{\phi} -\sqrt{{y}/{u}}\right)$}.

The nonlinear integral equation 
for $\Ls(r^{2})$
given by 
\1eq{lsg_renorm} is solved 
by employing the following sequence of steps: 
\begin{enumerate}[label=({\itshape\roman*})]
\item $\Delta(r^2)$, $F(r^2)$, and the ghost-gluon form factor $B_{1}(r^2, p^2, \theta)$ are treated as external inputs. For $\Delta(r^2)$ and $F(r^2)$ we use fits given by Eqs.~(C11) and~(C6) of~\cite{Aguilar:2021uwa}, respectively, to the lattice results of \cite{Aguilar:2021okw}, which have been cured from volume and discretization artifacts~\cite{Becirevic:1999uc,Becirevic:1999hj,deSoto:2007ht,deSoto:2022scb}. For $B_{1}(r^2, p^2, \theta)$ we employ the results of~\cite{Aguilar:2022thg,Ferreira:2023fva}
(see Figs.~13 and 14 of~\cite{Ferreira:2023fva}), which were obtained through the solution of the coupled system of SDEs for the ghost propagator and the ghost-gluon vertex, and reproduce the available lattice data of \cite{Ilgenfritz:2006he,Sternbeck:2006rd}.
All these inputs
are renormalized in the asymmetric MOM scheme, defined by \1eq{ren_conds}, at the renormalization point $\mu=4.3~\text{GeV}$. For this particular $\mu$ we use 
 \mbox{$\alpha_s = 0.27$}, as determined by the lattice simulation of~\cite{Boucaud:2017obn}.

\item To solve \1eq{lsg_renorm}, we first perform a change of variables
\be 
{\hat x} := \frac{x - 1}{ x + 1} \,, \label{change_var}
\ee
and similarly for all squared momenta appearing in \1eq{variosL}, including the integration measure. This procedure transforms the interval $[0,\infty]$ to the canonical interval $[-1,1]$. Evidently, we can rewrite any function of a squared momentum as a function of its hatted counterpart, \eg $\Ls(x)\to \Ls({\hat x})$,
and so on.

\item Then we expand $\Ls({\hat x})$ in terms of the Chebyshev polynomials of the second kind, $U_i({\hat x})$, \ie
\be 
\Ls({\hat x}) = \sum_{i = 0}^{N} c_i U_i({\hat x}) \,, \label{Ls_Cheby}
\ee
and similarly for $\Ls({\hat y})$ and $\Ls({\hat v})$, where we take $N = 39$. The resulting integrals are evaluated through Gauss–Legendre quadrature at 40 values of squared momenta in the range \mbox{$[8\times 10^{-4}, 1\times 10^{3}]~\text{GeV}^{2}$}.

\item At this point, \1eq{lsg_renorm} has been converted to a nonlinear algebraic system for the $40$ coefficients $c_i$,  which is solved using a quasi-Newton method. Finally, substituting the solution into \1eq{Ls_Cheby} and inverting \1eq{change_var} yields $\Ls(x)$.

\end{enumerate}

\begin{figure}[t]
 \includegraphics[width=0.45\linewidth]{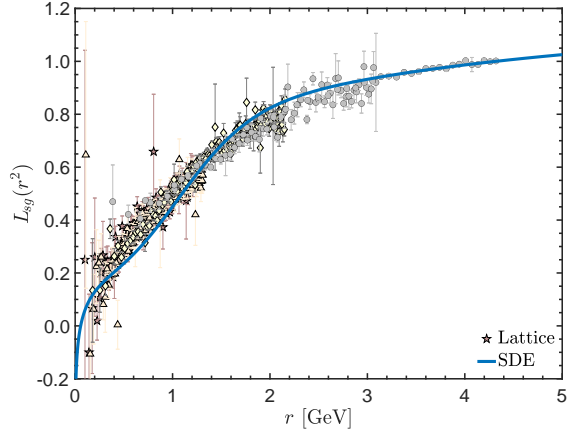}
 \caption{The  soft-gluon form factor $\Ls(r^{2})$ from~\cite{Aguilar:2021lke,Boucaud:2002fx,Boucaud:2003xi} (lattice data)  
compared with the SDE result obtained from \1eq{lsg_renorm} (blue continuous).}
\label{fig:lsg}
\end{figure}

The $\Ls(r^2)$ determined through the above procedure  is shown in \fig{fig:lsg} (blue continuous curve), where it is compared to the lattice data (points) of~\cite{Aguilar:2021lke,Boucaud:2002fx,Boucaud:2003xi}.  The observed agreement 
is particularly good: the deviation of the  
SDE solution from the lattice results is below $5\%$ for most of the momentum range 
where data exist.
The only exception is for $r\in[0.4,1.2]$~GeV, 
where the deviation is more pronounced; 
in particular, at momenta near $r = 0.6$~GeV the SDE result underestimates the lattice data  by $24\%$ at most.

\section{Infrared features revisited}\label{sec:suppression}

In this section we take advantage of the simplicity offered by \1eq{meq}
and revisit two particular features of $\Ls(r^2)$ 
in the Landau gauge, namely 
the infrared suppression of $\Ls(r^2)$ with respect to its tree-level value, and the divergence of this form factor at the origin. 
Even though both features have been extensively discussed in the recent literature 
at a qualitative level~\cite{Aguilar:2013vaa,Athenodorou:2016oyh,Boucaud:2017obn,Aguilar:2019uob,Aguilar:2021lke,Papavassiliou:2022umz,Ferreira:2023fva}, our results allow for 
a quantitative analysis of their origin. 

To this end, we begin by disentangling the contributions of the individual terms $d_{i,f}^{\scriptscriptstyle{sg}}(r^2,\mu^2)$ to the final result for $\Ls(r^2)$.  This is achieved by substituting the solution for $\Ls(r^{2})$, shown in \fig{fig:lsg}, into the expressions for the contributions of each diagram, given in \2eqs{lsg_renorm}{variosL}. The outcome of this exercise is presented on the left panel of \fig{fig:lsg},  where we
turn on, one by one, the resulting $d_{i,f}^{\scriptscriptstyle{sg}}(r^2,\mu^2)$. 

As we can see on the left panel of \fig{fig:lsg}, diagram $d_1$ has a positive contribution to $\Ls(r^2)$ in the infrared. On the other hand, both the ghost loops, $d_2$, and the swordfish diagrams, $d_3+d_4$, furnish a negative contribution in the infrared, thus suppressing $\Ls(r^2)$. However, it is clear that the bulk of the suppression has its origin in the swordfish terms. In fact, comparing the purple dotted and green dashed curves of \fig{fig:lsg}, corresponding to $1 + d_{1,f}^{\scriptscriptstyle{sg}}(r^2,\mu^2)$ and $1 + d_{1,f}^{\scriptscriptstyle{sg}}(r^2,\mu^2)+ d_{2,f}^{\scriptscriptstyle{sg}}(r^2,\mu^2)$, respectively, we see that the numerical impact of the ghost loop in $\Ls(r^2)$ is negligible for momenta $r \gtrapprox 0.3$~GeV. 
In other words, even if diagram $d_2$ were to be omitted entirely, 
the suppression of the vertex in the physically important region of momenta would remain practically unaltered. 

Next, we consider the behavior of $\Ls(r^2)$ near the origin. As has been shown in previous studies~\cite{Aguilar:2013vaa}, the nonperturbative masslessness of the ghost makes $d_{2,f}^{\scriptscriptstyle{sg}}$ diverge at $r = 0$. While this feature is already visible on the left panel of \fig{fig:lsg}, it is best appreciated in a logarithmic plot, presented on the right panel of the same figure. In this panel, we see that $\Ls(r^2)$ (blue continuous) displays a behavior consistent with a logarithmic divergence near the origin. To confirm that this divergence originates from $d_{2,f}^{\scriptscriptstyle{sg}}$, the result of $\Ls(r^2) - d_{2,f}^{\scriptscriptstyle{sg}}(r^2,\mu^2)$ is plotted in the same panel as a red dot-dashed curve, and clearly saturates to a constant value.

In previous studies~\cite{Aguilar:2013vaa,Aguilar:2021lke,Papavassiliou:2022umz}, the features of infrared suppression and divergence at the origin were thought to be connected:
the divergence of $d_{2,f}^{\scriptscriptstyle{sg}}(r^2,\mu^2)$ was 
understood to drive the suppression of $\Ls(r^2)$, since $d_{2,f}^{\scriptscriptstyle{sg}}(r^2,\mu^2)$ inevitably acquires values that are negative and large in magnitude for small enough $r$. Instead, from the results presented in \fig{fig:lsg}, it is clear that the infrared suppression of $\Ls(r^2)$ is driven by the swordfish diagrams, with the divergence of the ghost loop becoming apparent only at very small momenta.

\begin{figure}[t]
\includegraphics[width=0.45\linewidth]{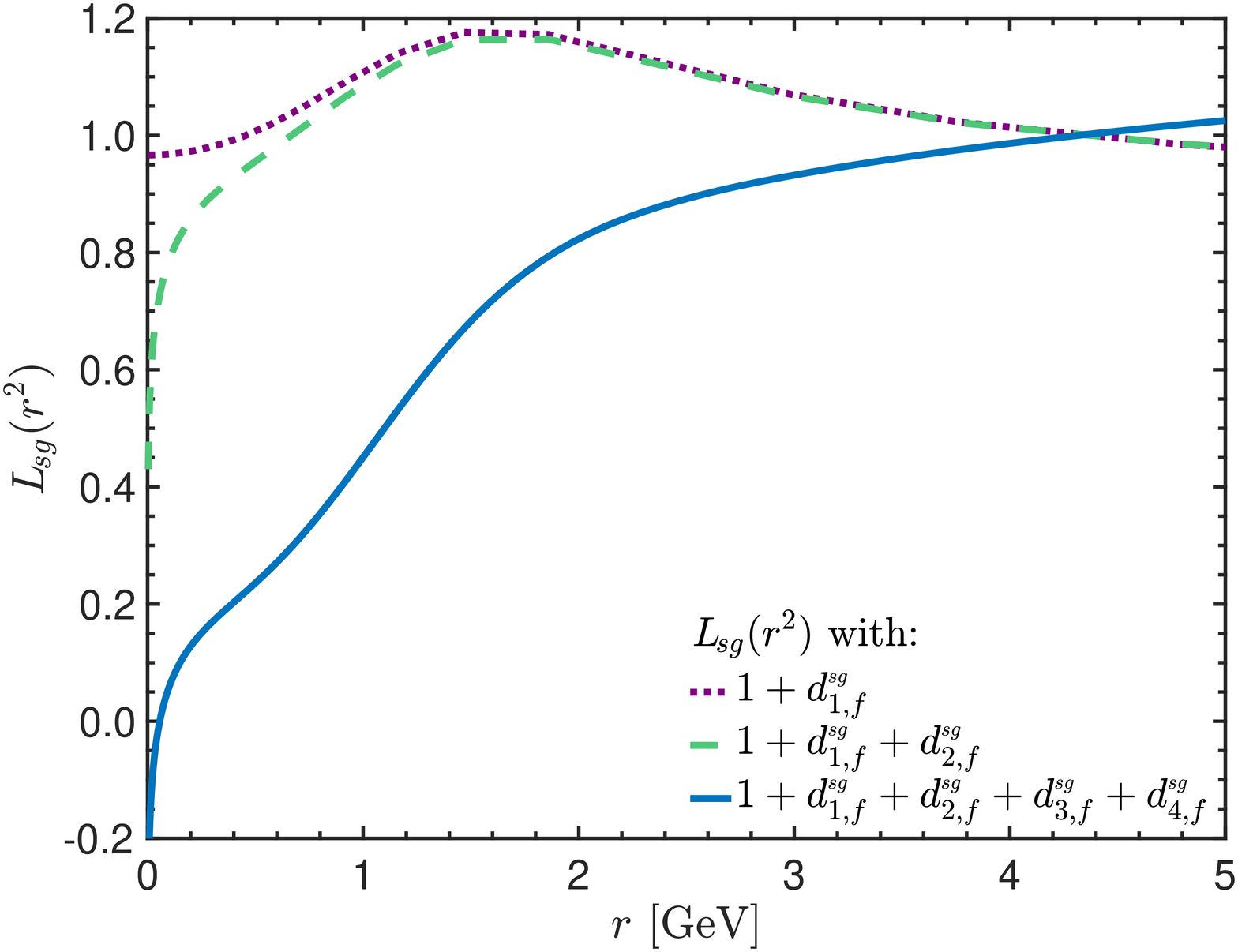} \hfil 
 \includegraphics[width=0.45\linewidth]{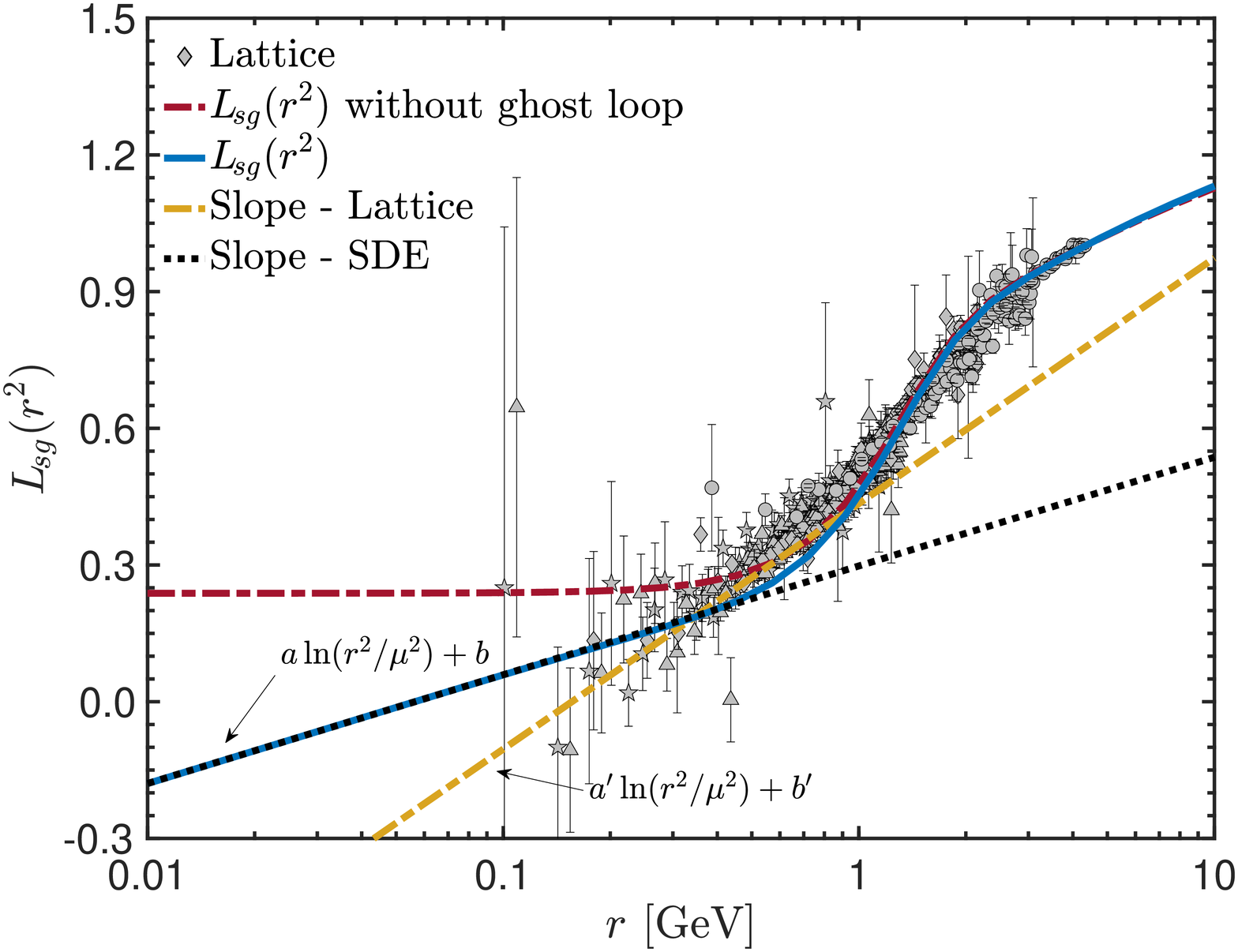}
 \caption{Left: The sequential inclusion of the diagrammatic contributions 
 $d_{i,f}^{\scriptscriptstyle{sg}}(r^2,\mu^2)$ 
that comprise the  soft-gluon form factor $\Ls(r^{2})$. 
 Right: Rate of the logarithmic divergence of $\Ls(r^{2})$ 
 extracted from ({\it i})  
 fitting the lattice data of~\cite{Aguilar:2021lke,Boucaud:2002fx,Boucaud:2003xi} (yellow dot-dashed curve),  
 and ({\it ii}) the SDE calculation leading to \1eq{asympt} (black dotted curve). }
\label{fig:lograte}
\end{figure}

Given these considerations, it is possible to obtain an exact expression describing the rate of the logarithmic divergence of $L_{sg}(r^{2})$ in the infrared by studying only the ghost diagram, $d_2$. This expression is derived in detail in the Appendix~\ref{App_divergence} and reads\footnote{This result was first derived as an all-order statement within the Curci-Ferrari model in~\cite{Barrios:2022hzr}.}
\be
L_{sg}(r^2) \underset{r^2\to 0}{\approx} = a\ln(r^2/\mu^2) + b \,, \qquad a = 
\left(\frac{\alpha_{s} C_{A}}{96\pi}\right){\widetilde Z}_1^{3}F^{3}(0) = 0.052\,, 
\label{asympt}
\ee
where ${\widetilde Z}_1(\mu = 4.3~\text{GeV})=0.933$ is the particular value assumed by the renormalization constant $Z_1$ of the ghost-gluon vertex in the asymmetric MOM scheme~\cite{Aguilar:2022thg,Ferreira:2023fva}. Lastly, $b$ is a finite constant left unspecified. 

On the right panel of \fig{fig:lograte}, we compare the SDE result for $\Ls(r^2)$ (blue continuous) to the asymptotic behavior given by \1eq{asympt} (black dotted), where we fixed $b=0.45$ by adjusting it to the full SDE solution. From this comparison, it is clear that the asymptotic behavior is only reached for momenta below $0.3$~GeV. For $r > 0.3$~GeV, the blue continuous and red dot-dashed curves become nearly equal, such that the slope of $\Ls(r^2)$ is clearly contaminated by positive contributions originating from the gluonic diagrams. 

The latter observation is relevant for the correct extraction of the asymptotic behavior of $\Ls(r^2)$ from the lattice data. Specifically, in~\cite{Aguilar:2021lke}, the parameters of \1eq{asympt} were determined by fitting \1eq{asympt} to the lattice data below $0.5$~GeV. This procedure yielded the values \mbox{$a\to a^{\prime}=0.117(6)$} and \mbox{$b\to b^{\prime}=0.78(8)$} (yellow dot-dashed line on the right panel of \fig{fig:lograte}). In particular, $a^{\prime}$ is $2.25$ times larger than the theoretical value found for $a$ in \1eq{asympt}.

Now, we note that out of the 52 data points of~\cite{Aguilar:2021lke,Boucaud:2002fx,Boucaud:2003xi} for $\Ls(r^2)$ with $r <0.5$~GeV, only 17 (about one-third) lie in the region $r < 0.3$~GeV, where $\Ls(r^2)$ is well-described by its asymptotic behavior. Moreover, these few points possess larger error bars than the points at higher momenta, thus having smaller weights in the fitting procedure. As such, the value of $a$ obtained through this procedure is polluted by non-asymptotic contributions. 

Indeed, if the fitting is repeated using only the 17 data points in the asymptotic region, \mbox{$r < 0.3$~GeV}, we obtain instead $a = 0.09(8)$. The latter number is consistent with the theoretical value of \1eq{asympt}, albeit with an error 
that is too large to be conclusive. 
Evidently, a more reliable comparison between SDE results and lattice would 
require data points for $r < 0.3$~GeV, which may be particularly costly. 

We conclude this section by emphasizing that the discussed features, and the 
diagrammatic origin attributed to them, have been explored in the strict confines 
of the Landau gauge; it would be interesting to explore 
if any of them persist for 
different values of the gauge-fixing parameter.

\section{General Kinematics}
\label{sec:genkin}

In this section we 
determine from the SDE the 
structure of the 
form factors $\gammanew_i(q^2,r^2,p^2)$ 
for general kinematics.
The main goal of this analysis is twofold: 
first, to establish quantitatively the 
extent of the dominance of the classical form factor, and 
second, to determine the accuracy of the planar degeneracy approximation, \ie of \1eq{planar_L}.

In general, the four form factors comprise a system of coupled integral equations, since all of the $\gammanew_i(q^2,r^2,p^2)$ contribute to the right-hand side of the SDE of \1eq{eq:finitedis}. To simplify our analysis, we will instead employ again the approximation given by \1eq{meq}, using for $\Ls(r^2)$ the result of the previous section, shown as a blue continuous line in \fig{fig:lsg}.

With the above procedure, the task of determining each $G_i(q^2,r^2,p^2)$ is reduced to the evaluation of a specific \emph{static} projection of the SDE of \1eq{eq:finitedis}. Concretely, the form factors can be extracted through
\be
\label{project}
\gammanew_{j}(q^2,r^2,p^2) = {\mathcal P}_{j}\cdot\overline{\fatg}(q,r,p)\,, \qquad {j=1,2,3,4\,,}
\ee
where  we have introduced the compact notation 
$A\cdot B := A^{\alpha\mu\nu}B_{\alpha\mu\nu}$. Evidently, the projectors ${\mathcal P}_{j}^{\alpha\mu\nu}$ can be defined by the requirement
\be 
{\mathcal P}_{j} \cdot \tau_{i} = \delta_{ij} \,. 
\label{P_def}
\ee

Since the basis elements, $\tau_i^{\alpha \mu \nu}$, defined in Eq.~\eqref{lambdaBasis}, are transversely-projected, the ${\mathcal P}_{j}^{\alpha\mu\nu}$ are themselves transverse, and may be expanded in the same basis, \ie
\begin{align}
{\mathcal P}_{j}^{\alpha\mu\nu} = \sum_{n}c_{jn} ~\tau_{n}^{\alpha \mu \nu} \,. 
\label{P_exp}
\end{align} 
Hence, combining \2eqs{P_def}{P_exp}, we obtain
\be 
\sum_{n}c_{jn} ~\tau_{n} \cdot\tau_{i} = \delta_{ij} \,.
\ee
The above expression can be conveniently expressed in matrix form by defining matrices $C$ and $T$ as
\be
[C]_{jn} := c_{jn} \,, \qquad [T]_{ni} := \tau_{n}\cdot\tau_{i} \,.
\ee
Then, \1eq{P_def} is rewritten  as
\be 
CT = \mathbb{1} \,, \quad \Longrightarrow \quad C = T^{-1} \,,
\label{inverse}
\ee
which provides a formal expression for the coefficients $c_{jn}$. We will not report here the expressions for the individual $c_{jn}$, 
since they are rather long\footnote{We point out that the resulting expressions for the ${\mathcal P}_{j}^{\alpha\mu\nu}$ are divergent whenever one momentum vanishes, or the magnitudes of two momenta are equal. Nevertheless, the form factors are finite in those limits, as can be shown by means of careful expansions of the projected integrals.}. Following the above steps, these expressions can easily be derived using any conventional program capable of performing Lorentz algebra, such as the \emph{Mathematica} packages \emph{Feyncalc}~\cite{Mertig:1990an} and \emph{Package-X}~\cite{Patel:2016fam}.   

Thus, in order to isolate the contributions of the form factors $\gammanew_{j}(q^2,r^2,p^2)$, defined in \1eq{project}, we
contract \1eq{eq:finitedis}  with the projectors of \1eq{P_exp}, which yield  (Minkowski space)
\begin{align}
{\gammanew}_{1}(q^{2},r^{2},p^{2}) &= 1 + \sum_{i=1}^{5} \mathcal{P}_1 \cdot \bar{d_i}(q,r,p)  - \sum_{i=1}^{4} d_i^{\,\scriptscriptstyle{sg}}(\mu^2) \,, \nonumber \\ 
{\gammanew}_{j}(q^{2},r^{2},p^{2}) &= \sum_{i=1}^{5} \mathcal{P}_j\cdot \bar{d_i}(q,r,p)\,, \qquad j=2,3,4\,,
\label{eq:2}
 \end{align}
where we have used the fact that $\tau_{1}^{\alpha \mu \nu} = \overline{\Gamma}_{\!0}^{\,\alpha \mu \nu}$
[see Eq.~\eqref{lambdaBasis}], and have applied   
the definition of \1eq{P_def} on the first term of Eq.~\eqref{eq:finitedis}.  

To proceed, we convert the expressions in \1eq{eq:2} to Euclidean space and use hyper-spherical coordinates. In doing so, the form factors are re-expressed as functions of $q^{2}$, $r^{2}$, and the angle between the four-vectors $q$ and $r$, $\theta_{qr}$, \ie we make the replacement \mbox{${\gammanew}_{i}(q^{2},r^{2},p^{2})\to {\gammanew}_{i}(q^{2},r^{2},\theta_{qr})$}. For the propagators, ghost-gluon vertex, and value of the coupling, we use the results described in item ({\it i}) of Sec.~\ref{sec:softgluon}.

Then, the numerical evaluation of the resulting expressions is performed on a grid of external momenta distributed logarithmically in the interval $q^2,\,r^2\in [10^{-3}, 10^{3}]\, \mbox{GeV}^2$ with 40 points in each dimension, 
while the angle $\theta_{qr}$ is uniformly distributed in the interval $[0, \pi]$ with $20$ points. It turns out that 
in certain kinematic regions, particularly near the soft-gluon limits, the triple integrations require multiple evaluations of the integrand in order to achieve acceptable precision, while, away from the soft-gluon limits, a few evaluations suffice. Thus, to perform the integration efficiently, we employ an adaptive quadrature method, namely the Gauss-Kronrod implementation of~\cite{Berntsen:1991:ADA:210232.210234}. Finally, all the needed interpolations in three variables are performed with B-splines \cite{de2001practical}.

\subsection{Form factor hierarchy}
\label{subsec:hierarchy}

We start our analysis of the general kinematics behavior of the form factors $\gammanew_i(q^2,r^2,\theta_{qr})$ by comparing their general forms and relative sizes.

\begin{figure}[t]
 \includegraphics[width=0.45\linewidth]{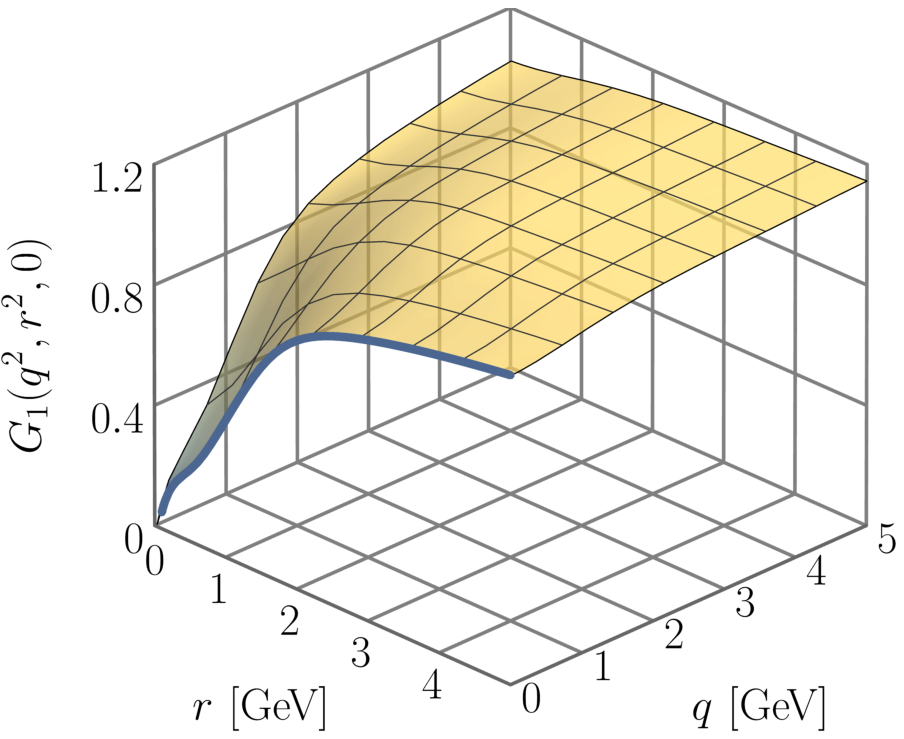}\hfil 
 \includegraphics[width=0.45\linewidth]{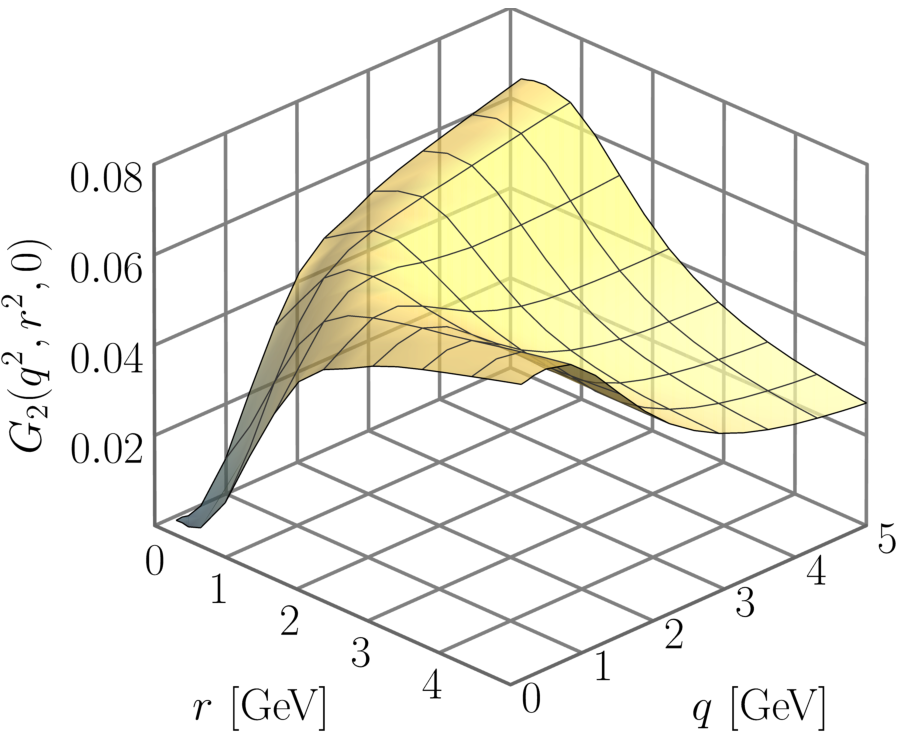}\\
 \includegraphics[width=0.45\linewidth]{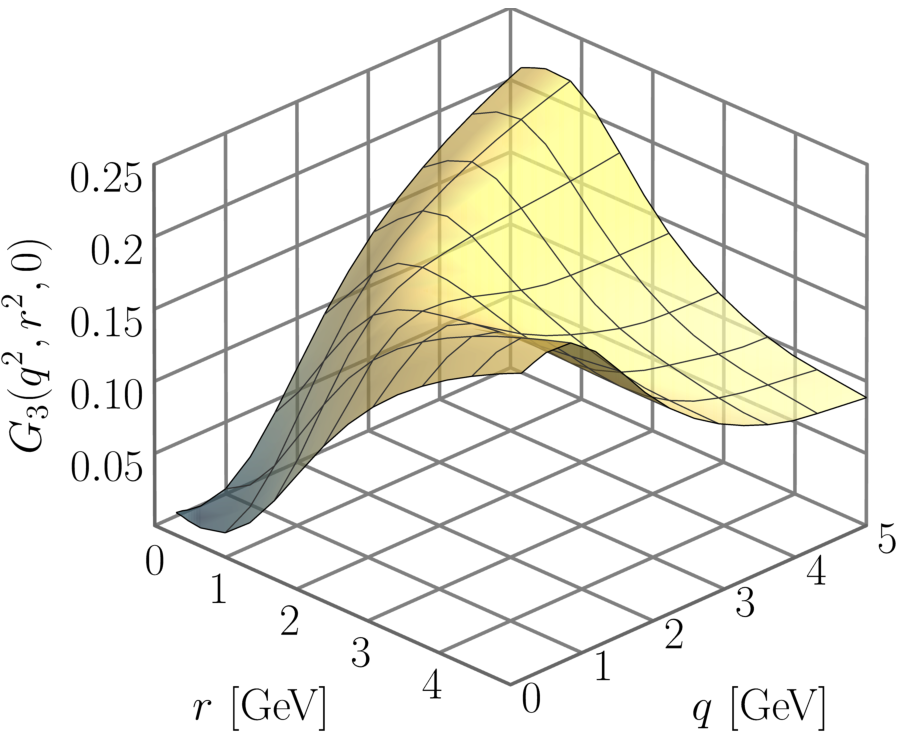}\hfil 
 \includegraphics[width=0.45\linewidth]{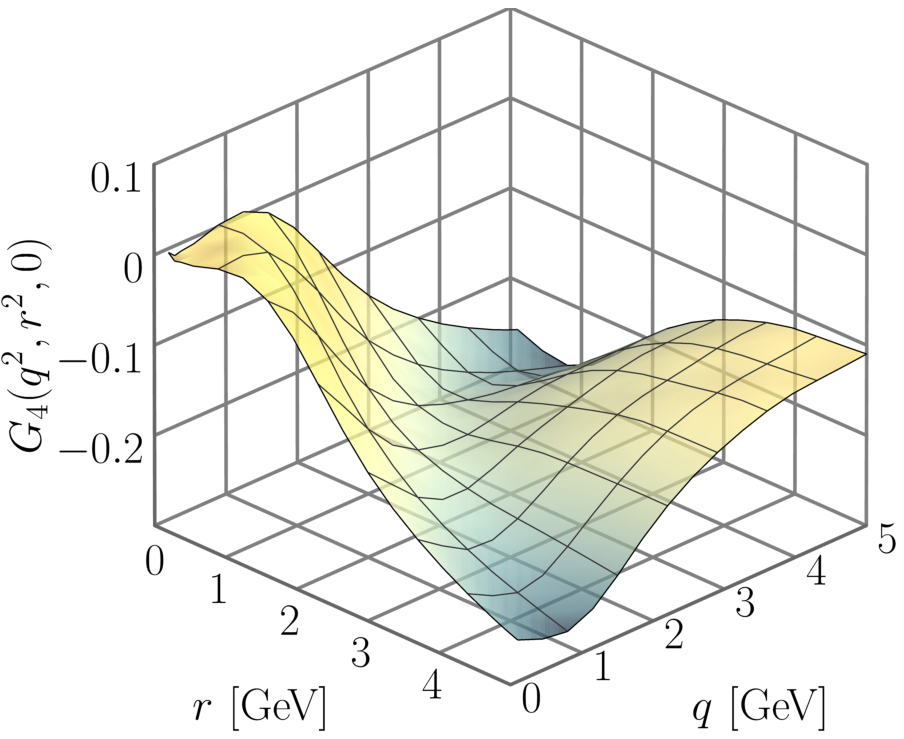}
 \caption{The vertex form factors $G_{i}(q^{2},r^{2},\theta_{qr})$, with $i= 1,2$ (top row) and $i=3,4$ (bottom row) plotted as functions of the magnitudes of the momenta $q$ and $r$, for the specific angle $\theta_{qr}=0$. In the first panel, we highlight (blue curve)
 the soft-gluon limit ($q=0$), corresponding to   
 the SDE solution for $L_{sg}(r^{2})$ displayed in \fig{fig:lsg}.}
 \label{fig:G1234}
\end{figure}

In \fig{fig:G1234} we show the $\gammanew_i(q^2,r^2,\theta_{qr})$ for the specific value of $\theta_{qr}=0$. For other values of $\theta_{qr}$, the results are qualitatively similar, with moderate quantitative differences. In the top left panel, we highlight as a blue solid curve the soft-gluon limit ($q=0$) of ${\gammanew}_{1}(q^{2},r^{2},\theta_{qr})$, for which we recover exactly the $L_{sg}(r^{2})$  of \fig{fig:lsg}, in agreement with \1eq{eq:gammaLsg}.

From \fig{fig:G1234}, we make the following observations:
\begin{enumerate}[label=({\itshape\roman*})]
\item First, only the classical form factor, $\gammanew_1(q^2,r^2,\theta_{qr})$, displays a divergence at the origin, namely the divergence discussed in Sec.~\ref{sec:suppression} and quantified by \1eq{asympt}. The remaining form factors are all found to be finite at the origin.

\item Next, we note that the form factors $\gammanew_{2,3,4}(q^2,r^2,\theta_{qr})$ are subleading in comparison to $\gammanew_1(q^2,r^2,\theta_{qr})$, 
for most of the kinematic range. In particular, $\gammanew_2(q^2,r^2,\theta_{qr}$) is found to be the smallest of all, and positive through the entire range, while 
$\gammanew_3(q^2,r^2,\theta_{qr}$) and $\gammanew_4(q^2,r^2,\theta_{qr}$) have comparable magnitudes and opposite signs.

\item In the soft-gluon limit, $q = 0$ (as well as $r = 0$ and $p = 0$, by Bose symmetry), the magnitudes of the form factors $\gammanew_{2,3,4}(0,r^2,\theta_{qr})$ become increasing functions of the remaining momentum, $r$. Therefore, for 
sufficiently large $r$, the $\gammanew_{2,3,4}(0,r^2,\theta_{qr})$ 
become comparable in magnitude to $\gammanew_{1}(0,r^2,\theta_{qr})$. 
In particular, at $r = 5$~GeV the $\gammanew_3$ and $\gammanew_4$ reach values of $0.21$ and $-0.25$, respectively, which correspond to $20\%$ and $-25\%$ of the value of $\gammanew_1$ at the same point. For larger $r$, these proportions increase further, such that $\gammanew_{2,3,4}(0,r^2,\theta_{qr})$ become comparable in magnitude to the classical form factor. 
As has been discussed in~\cite{Pinto-Gomez:2022brg} (see Sec.~6 and Fig.~7 therein),
the enhancement of the $\gammanew_{2,3,4}(0,r^2,\theta_{qr})$ 
may be interpreted as a finite remainder of a 
would-be collinear divergence, averted by the emergence of the 
nonperturbative gluon mass.
Note finally that the above behavior 
does not invalidate the approximation given by \1eq{meq}, because, as $q\to 0$,
the associated basis elements $\tau_{2,3,4}^{\alpha\mu\nu}$  
vanish linearly in $q$.

\end{enumerate}

\subsection{Planar degeneracy}\label{subsec:planardeg}

We next analyze the accuracy of the planar degeneracy approximation, \ie \1eq{planar}, and consider whether this property can be generalized to the subleading form factors.

\begin{figure}[t]
 \includegraphics[width=0.45\linewidth, trim={0 0 0 0.5cm},clip]{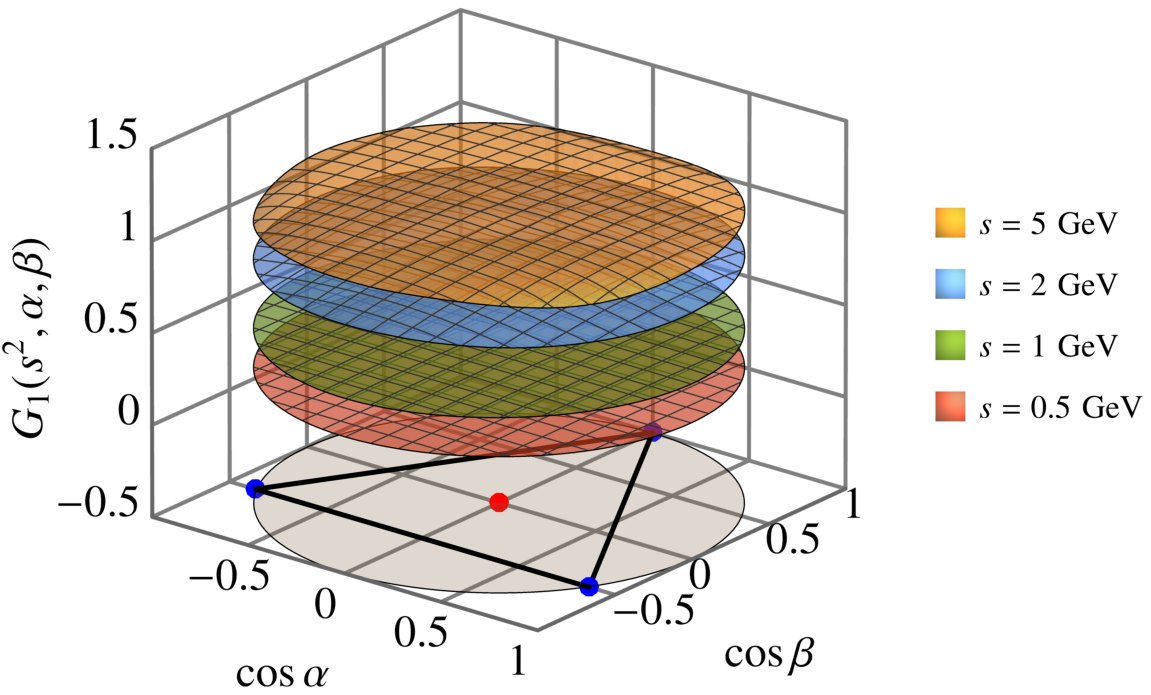}\hfil 
 \includegraphics[width=0.45\linewidth, trim={0 0 0 0.5cm},clip]{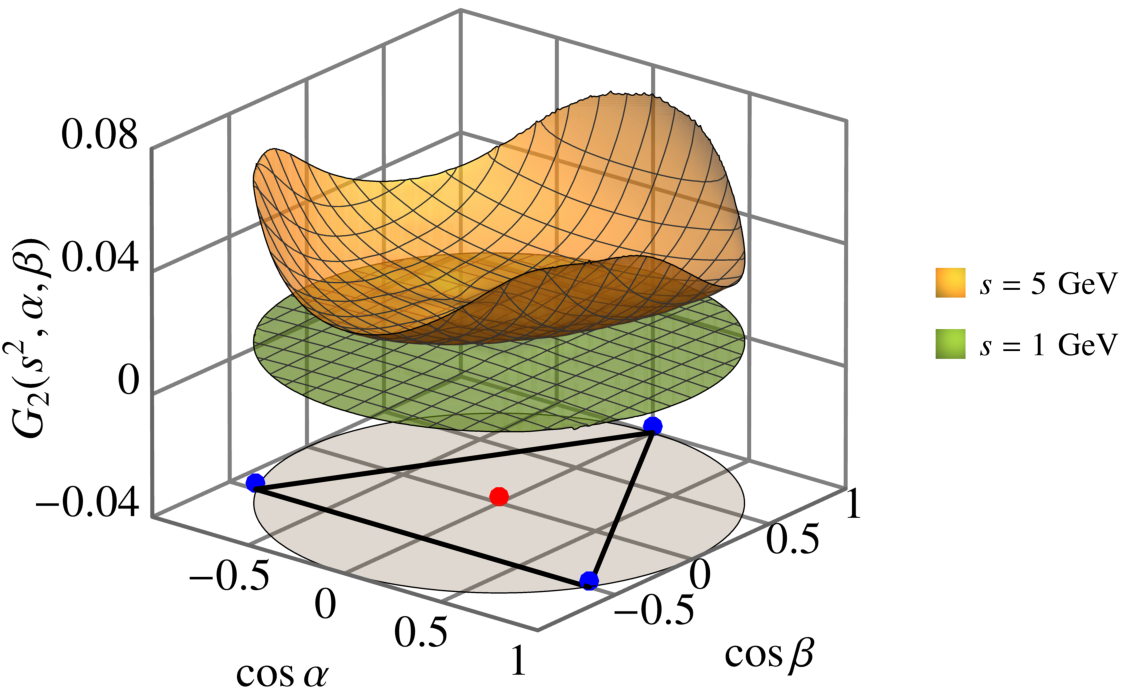}\\
 \includegraphics[width=0.45\linewidth, trim={0 0 0 0.5cm},clip]{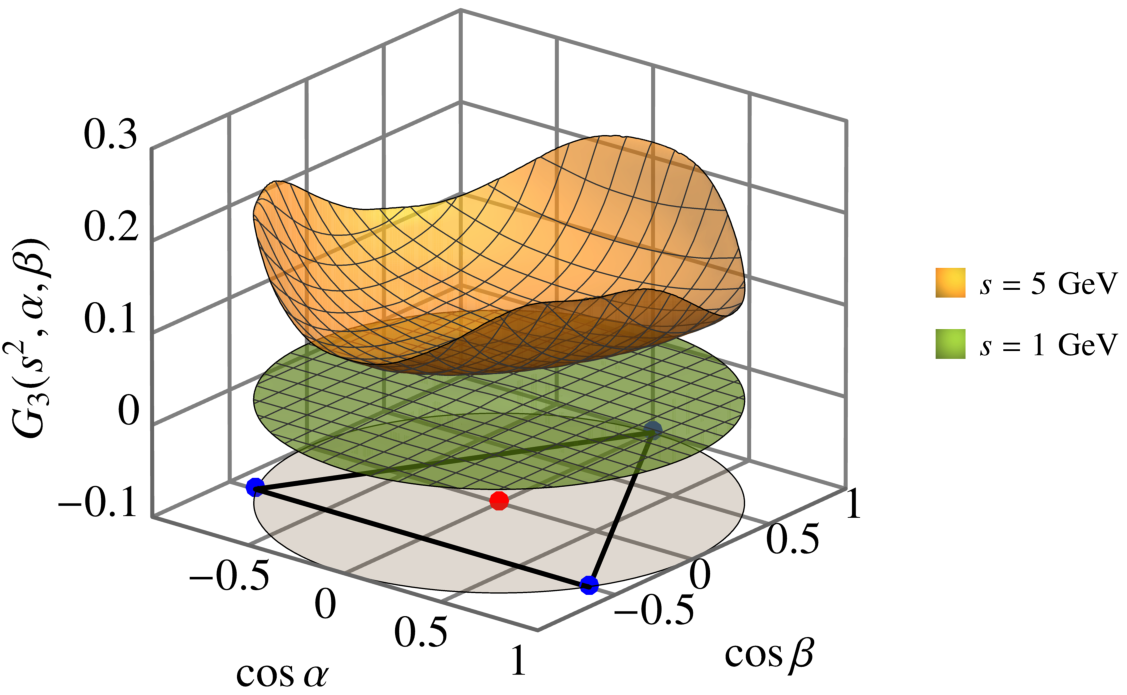}\hfil 
 \includegraphics[width=0.45\linewidth, trim={0 0 0 0.5cm},clip]{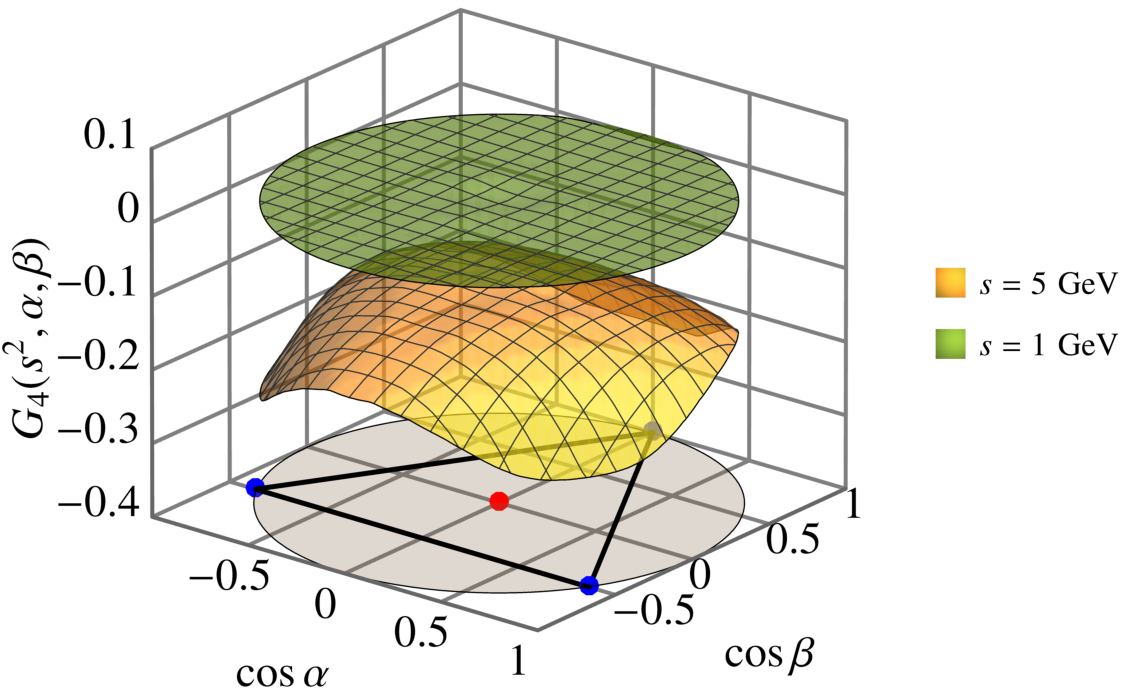}
 \caption{ The form factors $G_i(s^2,\alpha,\beta)$, shown as functions of the variables $s^2$, $\alpha$ and $\beta$, defined in \2eqs{s_variable}{Pgroup_variables}. }
 \label{fig:G1234_planar}
\end{figure}

To that end, it is convenient to reparametrize the form factors $\gammanew_i$ in terms of the variable $s^2$ of \1eq{s_variable}, rather than the individual momenta $q$, $r$ and $p$. This can be achieved by introducing two new angles, $\alpha$ and $\beta$, defined by\footnote{The variables $(s^2,\cos\alpha,\cos\beta)$ correspond to $(3{\cal S}_0,a,s)$ in the notation of \cite{Eichmann:2014xya} and to the variables $(s^2,2b/\sqrt{3},-6a/\sqrt{3})$ of \cite{Pinto-Gomez:2022brg}.}
\be 
\cos\alpha := \frac{\sqrt{3}(r^2 - q^2)}{2 s^2} \,, \qquad \cos\beta := \frac{q^2 + r^2 - 2 p^2}{2 s^2} \,. \label{Pgroup_variables}
\ee
Note that momentum conservation implies
\be 
\cos^2\alpha + \cos^2\beta \leq 1 \,,
\ee
\ie the kinematically allowed range of $\cos\alpha$ and $\cos \beta$ is the unit disk, which is represented in gray in each of the panels of \fig{fig:G1234_planar}.

The form factors can then be expressed as $G_i(s^2,\alpha,\beta)$ through the inverse relations
\begin{align}
q^2 =& \frac{s^2}{3}(2 + \cos\beta - \sqrt{3}\cos\alpha ) \,, \nonumber\\
r^2 =& \frac{s^2}{3}(2 + \cos\beta + \sqrt{3}\cos\alpha )  \,, \nonumber\\
\cos\theta_{qr} =& - \frac{1 + 2 \cos\beta}{\sqrt{(2 + \cos\beta)^2 - 3 \cos^2\alpha}} \,.
\end{align}
To facilitate the comparison between the different parametrizations of $G_i$, we list below how some special kinematic limits are represented in the $(s^2,\alpha,\beta)$ coordinate system:
\begin{enumerate}[label=({\itshape\roman*})]
\item The \emph{totally symmetric limit}, $q^2 = r^2 = p^2$, corresponds to the center of the disk, \ie \mbox{$\cos \alpha =\cos\beta = 0$}, and is represented by a red dot in each of the panels of \fig{fig:G1234_planar}.

\item The soft-gluon limit, $q = 0$, and its Bose symmetric counterparts $r = 0$ and $p = 0$ are given by
\begin{align}
q = 0 \quad \Leftrightarrow&\, \qquad  (\cos\alpha\,, \cos\beta) = \left(\sqrt{3}/2,-1/2\right)\,, \nonumber\\
r = 0 \quad \Leftrightarrow&\, \qquad  (\cos\alpha\,, \cos\beta) = - \left(\sqrt{3}/2,1/2\right)\,, \nonumber\\
p = 0 \quad \Leftrightarrow&\, \qquad  (\cos\alpha\,, \cos\beta) = \left(0,1\right)\,,
\end{align}
which correspond to the vertices of an equilateral triangle inscribed in the unit circle. These points and the triangle they form are represented by blue dots and black lines, respectively, in each of the panels of \fig{fig:G1234_planar}.

\end{enumerate}
%

The $G_i(s^2,\alpha,\beta)$ are shown in \fig{fig:G1234_planar} for general values of $\alpha$ and $\beta$ and selected values of $s$. Specifically, for the case of 
$G_1(s^2,\alpha,\beta)$, 
we show surfaces corresponding to 
\mbox{$s = 0.5 $~GeV} (red),  
\mbox{$s = 1 $~GeV} (green), 
\mbox{$s = 2 $~GeV}  (blue), and 
\mbox{$s = 5$~GeV} (yellow). 
In the case of 
the subleading form factors, the dense overlap of the resulting surfaces 
makes their visual distinction difficult; we therefore show only 
two examples, \mbox{$s = 1 $~GeV} and \mbox{$s = 5$~GeV}.

On the top left panel of \fig{fig:G1234_planar}, we see clearly that for fixed $s$ the classical form factor is rather flat, \ie nearly independent of $\alpha$ and $\beta$. Hence, the approximate planar degeneracy property of \1eq{planar} is verified
at a high level of accuracy. 

As for the subleading form factors, $G_{2,3,4}(s^2,\alpha,\beta)$, we note that the corresponding surfaces in \fig{fig:G1234_planar} are flat for $s = 1$~GeV, but not so for $s = 5$~GeV. Instead, in the latter case, the corresponding surfaces
increase markedly when one of the momenta vanishes. Evidently, this effect 
corresponds to the increase in the magnitudes of the $G_{2,3,4}$ 
near the soft-gluon limit, already discussed in relation with \fig{fig:G1234}. 
From this analysis, we conclude that planar degeneracy would be a poor approximation for the subleading form factors, except at \mbox{$s\lessapprox1$~GeV}.

Returning to the classical form factor, it is important to quantify the accuracy of the planar degeneracy approximation. To this end, we define the function
\be 
d(q^2,r^2,\theta_{qr}) := \frac{G_1(q^2,r^2,\theta_{qr}) - \Ls(s^2)}{G_1(q^2,r^2,\theta_{qr})}\times 100 \% \,, \label{d_err_def}
\ee
which measures in percentages the error made in approximating $G_1(q^2,r^2,\theta_{qr})$ by $\Ls(s^2)$. Note that, recalling \1eq{eq:gammaLsg}, $d(q^2,r^2,\theta_{qr})$ vanishes whenever one momentum is zero.

\begin{figure}[t]
 \includegraphics[width=0.31\linewidth]{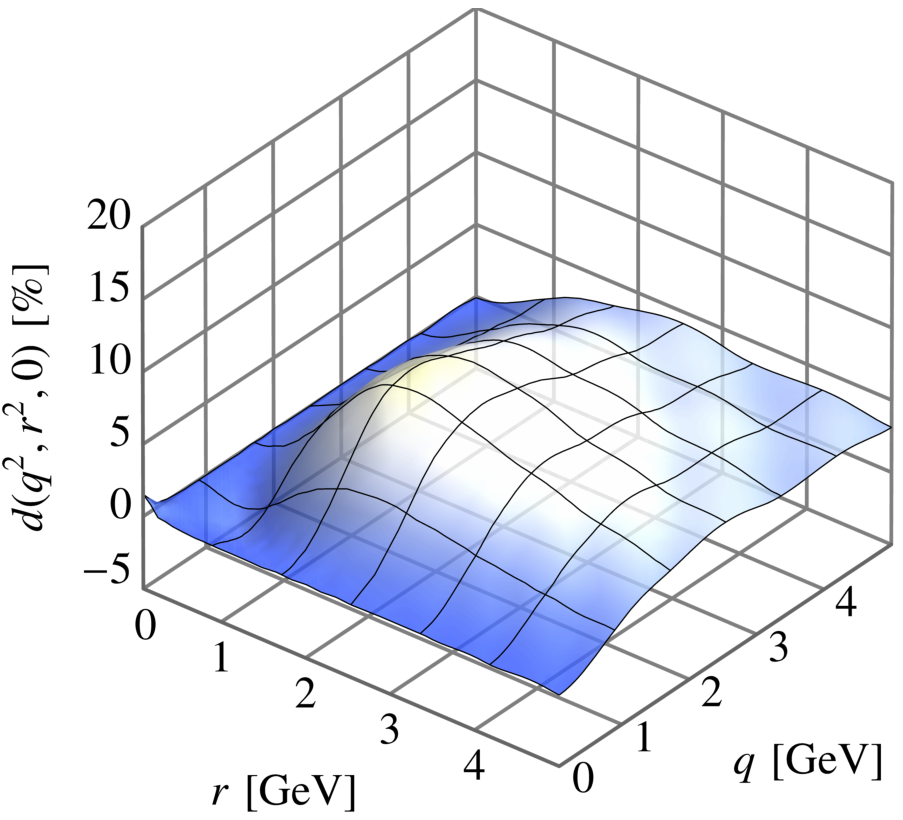}\hfil 
 \includegraphics[width=0.31\linewidth]{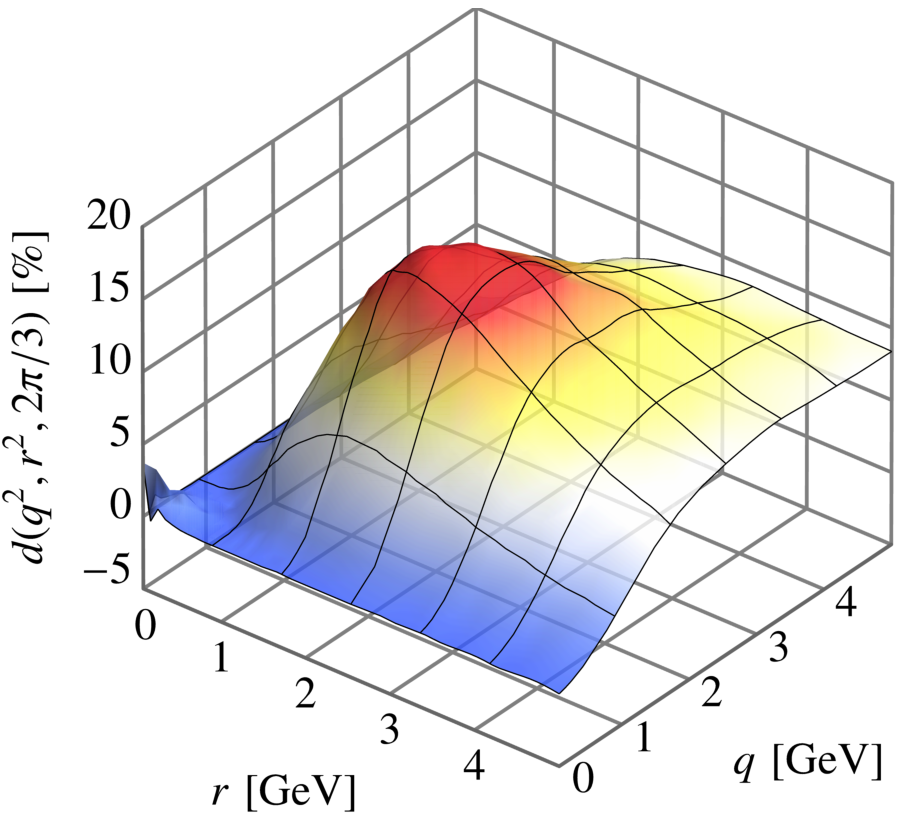}\hfil
  \includegraphics[width=0.31\linewidth]{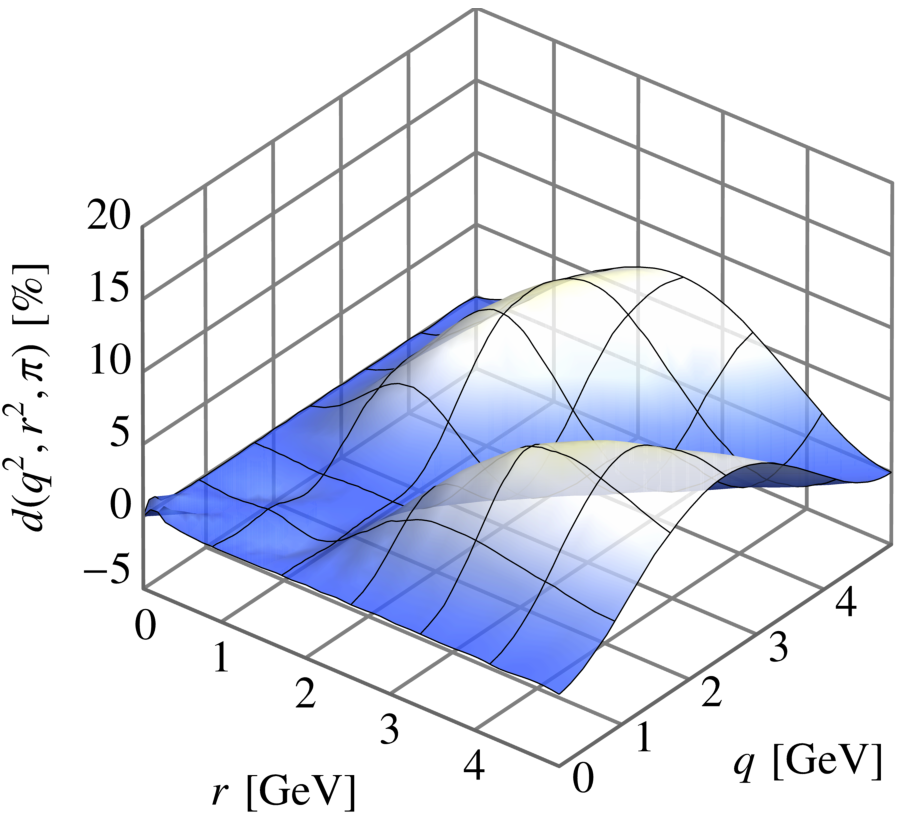}
 \caption{ Error measure, $d(q^2,r^2,\theta_{qr})$, defined in \1eq{d_err_def}, for $\theta_{qr} = 0$ (left), $\theta_{qr} = 2\pi/3$ (center), and $\theta_{qr} = \pi$ (right). }
 \label{fig:G1_planar_err}
\end{figure}

The maximum error can be found using standard numerical methods and is given by \mbox{$d_{\srm{max}} = 17.5\%$}. This value of error is attained at the symmetric point \mbox{$q = r = p = 2.0$~GeV}, corresponding to \mbox{$\theta_{qr} = 2\pi/3$}. For angles away from $\theta_{qr} = 2\pi/3$, the error diminishes quickly, falling below $10\%$ for most of the range. This can be seen clearly in \fig{fig:G1_planar_err}, where $d(q^2,r^2,\theta_{qr})$ is plotted for \mbox{$\theta_{qr} = 0$}, \mbox{$\theta_{qr} = 2\pi/3$} and  \mbox{$\theta_{qr} = \pi$}.

\begin{figure}[t]
 \includegraphics[width=0.4\linewidth]{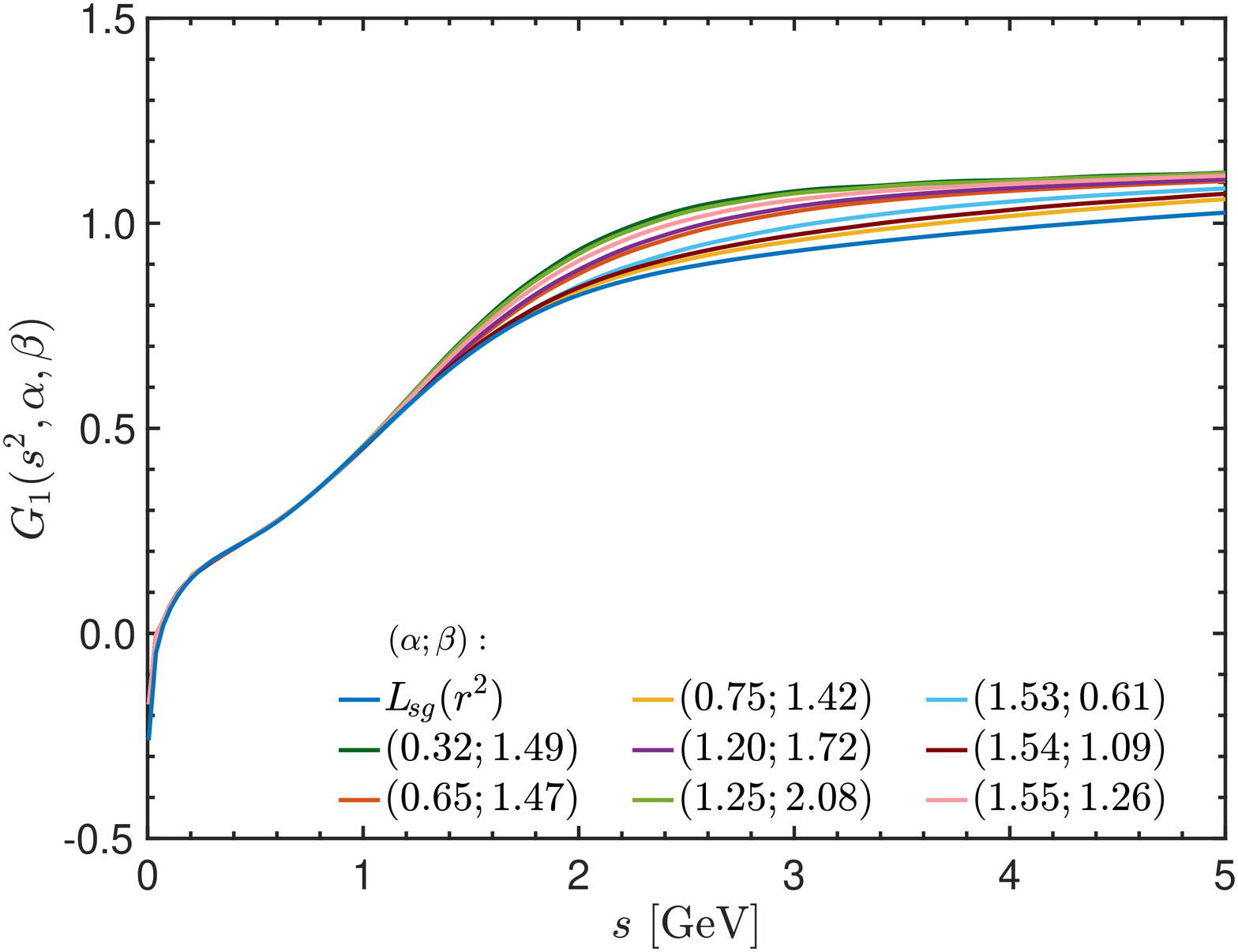}\hfil \quad 
 \includegraphics[width=0.50\linewidth]{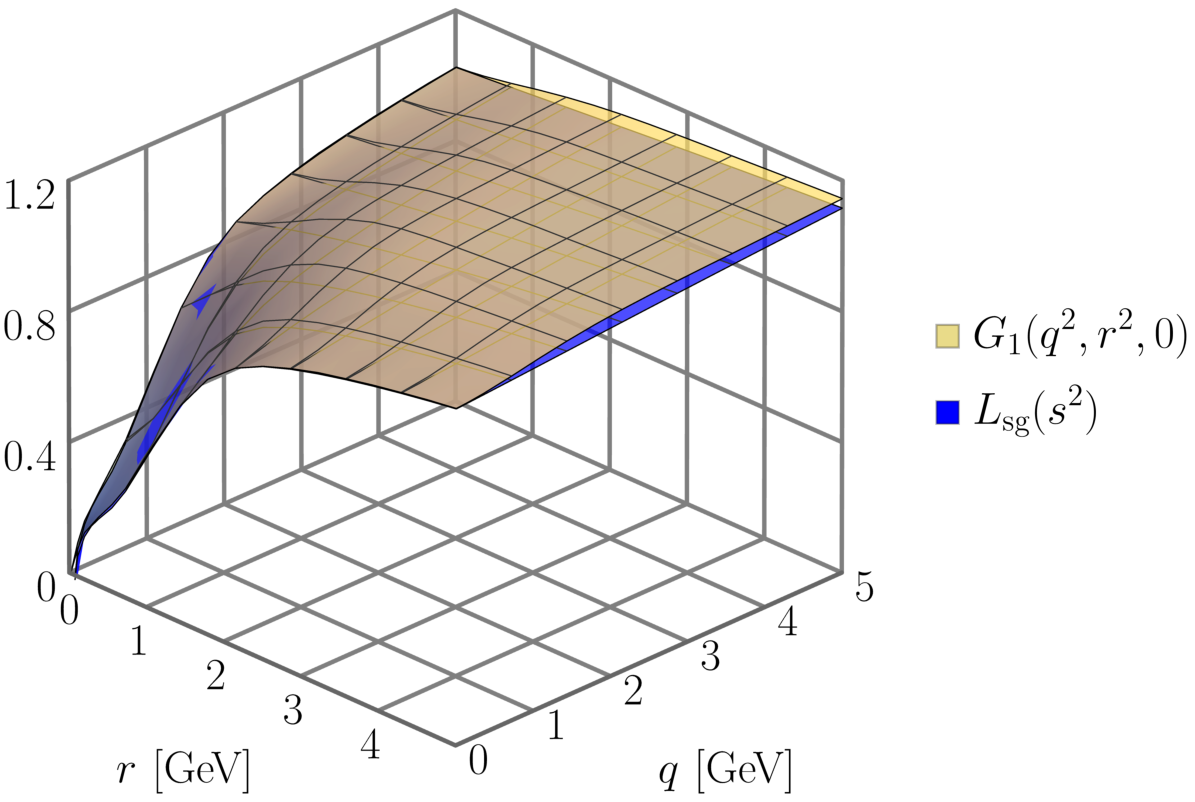}
 \caption{Left: Comparison of $G_1(s^2,\alpha,\beta)$,
for 8 randomly chosen values of $\alpha$ and $\beta$, to the soft-gluon limit $\Ls(s^2)$. The values of $\alpha$ and $\beta$ are given in radians in the legend. Right: Comparison between $G_1(q^2,r^2,\theta_{qr})$ and $\Ls(s^{2})$, where $s^{2} = q^{2} + r^{2} + qr\cos\theta_{qr}$, for arbitrary $q^{2}$ and $r^{2}$, and the choice of angle $\theta_{qr} = 0$. Note that in the soft-gluon limit ($q\to 0$) both surfaces become equal; the resulting curve corresponds to $\Ls(r^2)$.}
 \label{fig:G1_vs_Lsg}
\end{figure}

To conclude, we note that the error $d(q^2,r^2,\theta_{qr})$ in the planar degeneracy approximation is positive, apart from minor fluctuations in the deep infrared, as seen in \fig{fig:G1_planar_err}. Hence, \1eq{planar_L} tends to \emph{underestimate} the value of $G_1(q^2,r^2,\theta_{qr})$. This is evident on the left panel of \fig{fig:G1_vs_Lsg}, where  $G_1(s^2,\alpha,\beta)$ is compared to the soft-gluon limit, $\Ls(s^2)$, 
for 8 randomly chosen values of $\alpha$ and $\beta$.
Evidently, in all cases, when \mbox{$s \gtrapprox 1.4$~GeV}, the form factor $G_1$ exceeds $\Ls(s^2)$.  The same behavior can be appreciated more generally on the right panel of \fig{fig:G1_vs_Lsg}, where $G_1(q^2,r^2,\theta_{qr})$ (yellow) is seen to be above $\Ls(s^2)$ (blue), for the representative angle \mbox{$\theta_{qr} = 0$} and values of $q^2$ and $r^2$ larger than about $1$~GeV.

\subsection{Totally symmetric limit}\label{subsec:sym}

Our final exercise is to 
specialize our results to the \emph{totally symmetric} configuration, defined by the condition \mbox{$q^2 = r^2 = p^2 := Q^2$}, where $Q^2$ denotes the single momentum scale available. Note that this condition implies \mbox{$q\cdot r = r\cdot p = p\cdot q = - Q^2/2$}, and \mbox{$\theta_{qr} = \theta_{rp} = \theta_{pq} = 2\pi/3$}. 

In this limit, the tensors $\tau_i^{\alpha\mu\nu}(q,r,p)$ of \1eq{lambdaBasis} become linearly dependent, such that the tensor decomposition of $\overline{\fatg}^{\,\alpha\mu\nu}(q,r,p)$ collapses to~\cite{Athenodorou:2016oyh,Boucaud:2017obn,Aguilar:2019uob,Aguilar:2021lke}
\be
\overline{\fatg}^{\,\alpha\mu\nu}(q,r,p) = {\overline \Gamma}_1^{\,\srm{sym}}(Q^2) \lambda_1^{\alpha\mu\nu}(q,r,p) + {\overline \Gamma}_2^{\,\srm{sym}}(Q^2) \lambda_2^{\alpha\mu\nu}(q,r,p) \,,
\ee
where
\be 
\lambda_1^{\alpha\mu\nu}(q,r,p) := \overline{\g}_{\!0}^{\,\alpha \mu \nu}(q,r,p) \,, \qquad \lambda_2^{\alpha\mu\nu}(q,r,p) := \frac{(q-r)^{\nu} (r-p)^{\alpha} (p-q)^{\mu}}{Q^2} \,.
\ee
Then, it is straightforward to show that the form factor ${\overline \Gamma}_1^{\,\srm{sym}}(Q^2)$, associated with the tree-level tensor structure, can be obtained through the projection~\cite{Athenodorou:2016oyh,Boucaud:2017obn,Aguilar:2019uob,Aguilar:2021lke}
\be 
{\overline \Gamma}_1^{\,\srm{sym}}(Q^2) = \frac{W^{\alpha\mu\nu}(q,r,p)\overline{\fatg}^{\,\alpha\mu\nu}(q,r,p)}{W^{\alpha\mu\nu}(q,r,p)W_{\alpha\mu\nu}(q,r,p)}\rule[0cm]{0cm}{0.5cm} \Bigg|_{q^2=r^2=p^2=Q^2} \,, \label{Gamma_1_sym_proj}
\ee
where
\be 
W^{\alpha\mu\nu}(q,r,p) := \lambda_1^{\alpha\mu\nu}(q,r,p) + \frac{1}{2}\lambda_2^{\alpha\mu\nu}(q,r,p) \,.
\ee

\begin{figure}[t]
 \includegraphics[width=0.45\linewidth]{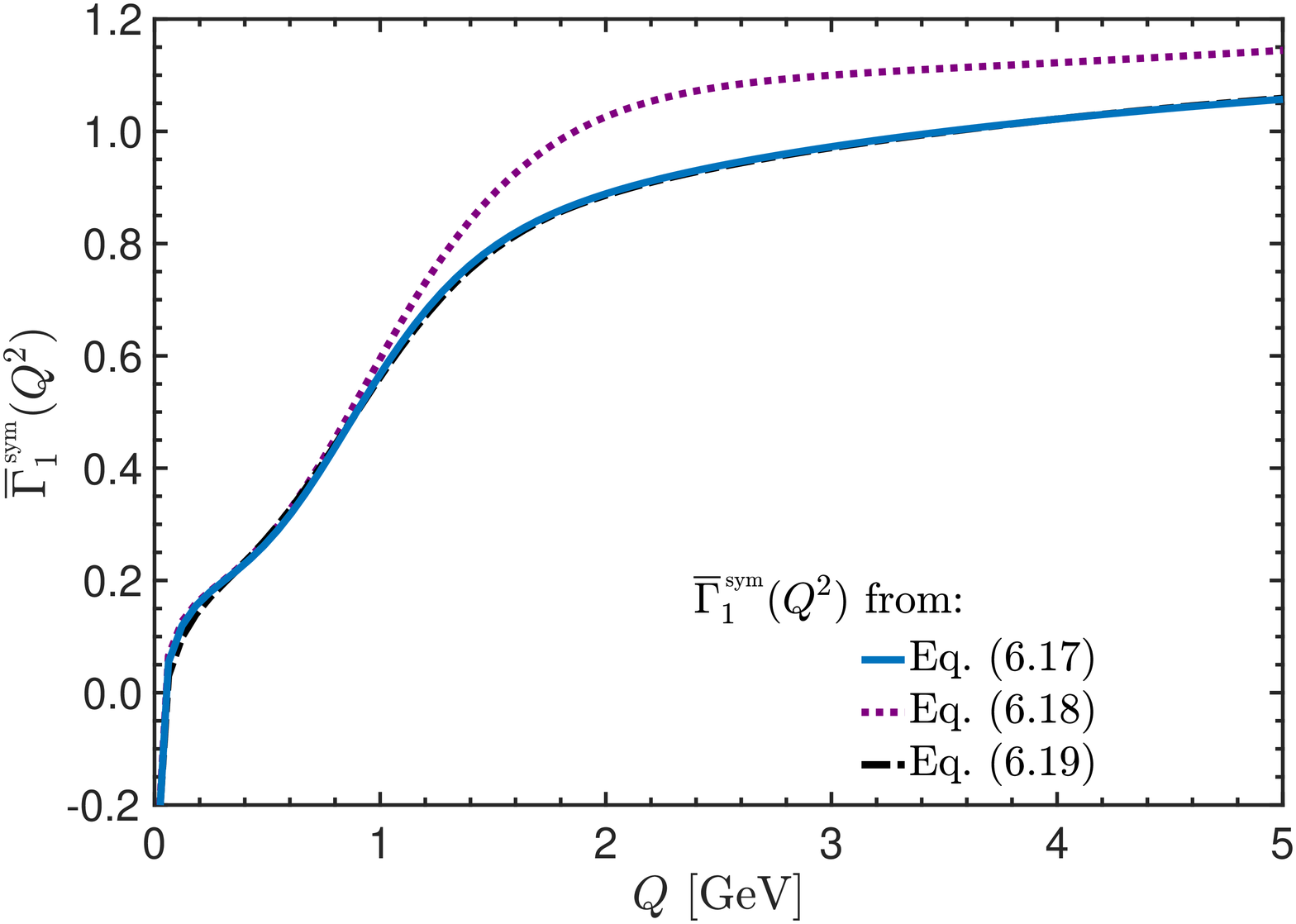}\hfil \quad 
 \includegraphics[width=0.45\linewidth]{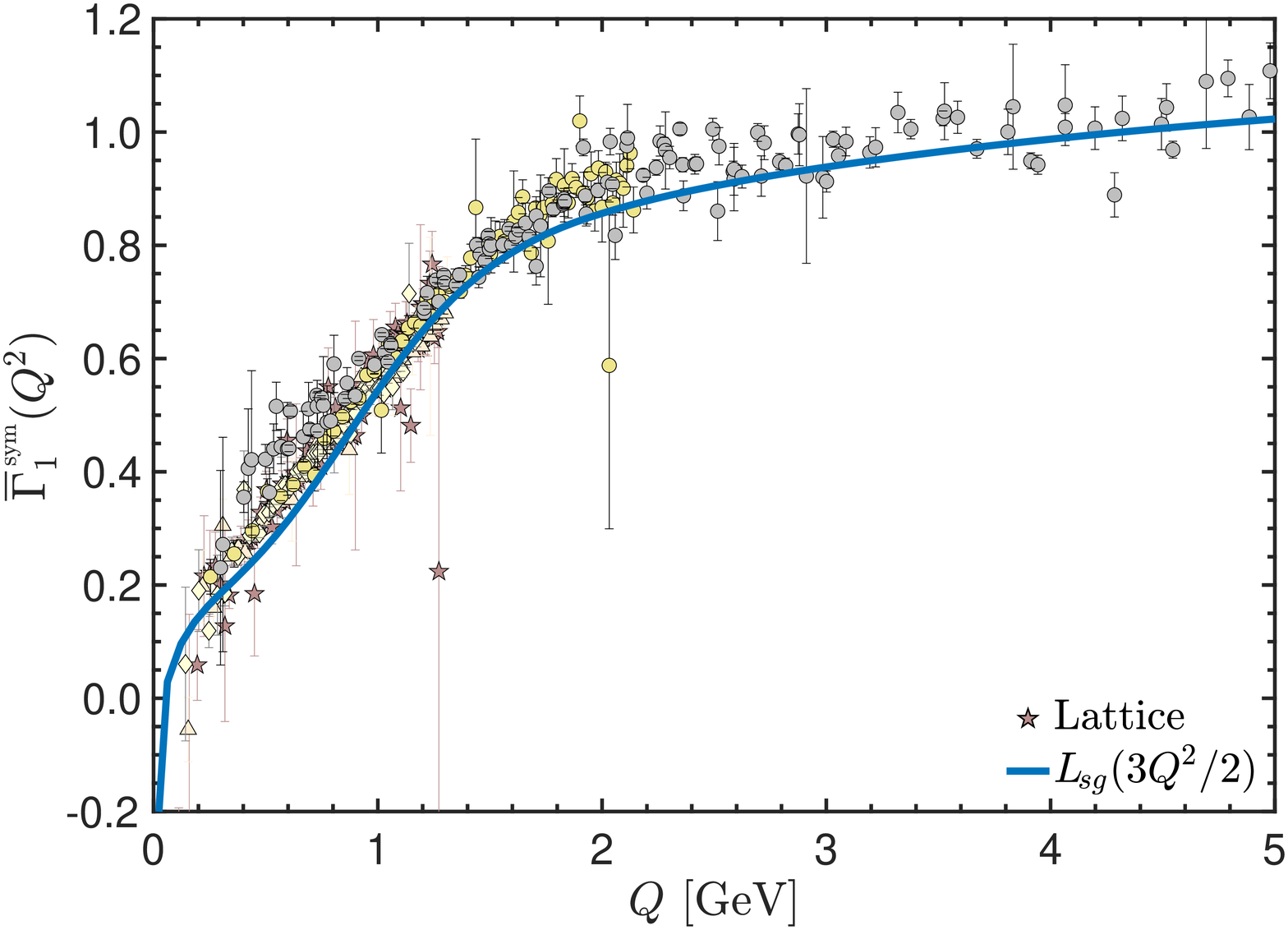}
 \caption{ Left: SDE result for ${\overline \Gamma}_1^{\,\srm{sym}}(Q^2)$ obtained from \1eq{Gamma_1_sym_from_Gi} (blue continuous), compared to the two approximations given by \2eqs{Gamma_1_sym_from_G1}{Gamma_1_sym_from_Lsg} (purple dotted  and black dot-dashed, respectively). Right: Lattice data of~\cite{Aguilar:2021lke,Boucaud:2002fx,Boucaud:2003xi} for ${\overline \Gamma}_1^{\,\srm{sym}}(Q^2)$, compared to the SDE result combined with \1eq{Gamma_1_sym_from_Lsg} (blue continuous line).}
 \label{fig:Gamma_1_sym}
\end{figure}

The \emph{exact} correspondence between the form factors ${\overline \Gamma}_1^{\,\srm{sym}}$ and $G_i$ is obtained by using  \1eq{Gamma_1_sym_proj} in \1eq{new_projection2}; it reads
\be 
{\overline \Gamma}_1^{\,\srm{sym}}(Q^2) = G_1(Q^2) - G_3(Q^2) + G_4(Q^2)/4 \,, \label{Gamma_1_sym_from_Gi}
\ee
where we write \mbox{$G_i(Q^2,Q^2,Q^2)\to G_i(Q^2)$} for the symmetric limits of the form factors. Using for the $G_i$ the results
 discussed in the previous subsections, we obtain for ${\overline \Gamma}_1^{\,\srm{sym}}(Q^2)$ the blue continuous line shown on the left panel of \fig{fig:Gamma_1_sym}.

Then we consider the effect of neglecting the subleading form factors, $G_3$ and $G_4$, in \1eq{Gamma_1_sym_from_Gi}. In this case, we obtain the approximation
\be 
{\overline \Gamma}_1^{\,\srm{sym}}(Q^2) \approx G_1(Q^2) \,, \label{Gamma_1_sym_from_G1}
\ee
which is shown as a purple dotted line on the left panel of \fig{fig:Gamma_1_sym}.

Lastly, we consider the prediction of the compact expression \1eq{meq}, noting that in the symmetric limit $s^2\to 3Q^2/2$. Evidently, in this case
\be 
{\overline \Gamma}_1^{\,\srm{sym}}(Q^2) \approx \Ls(3Q^2/2) \,, \label{Gamma_1_sym_from_Lsg}
\ee
which leads to the result displayed as a black dot-dashed line on the left panel of \fig{fig:Gamma_1_sym}.

From the results shown on the left panel of \fig{fig:Gamma_1_sym}, we see that the approximation in \1eq{Gamma_1_sym_from_G1}
(purple dotted) overestimates the true value of ${\overline \Gamma}_1^{\,\srm{sym}}(Q^2)$ (blue continuous), given by \1eq{Gamma_1_sym_from_Gi}.

Quite remarkably, the approximation given in \1eq{Gamma_1_sym_from_Lsg}
(black dot-dashed), derived from \1eq{meq}, reproduces the full result to within $3\%$.  The exceptional accuracy of \1eq{Gamma_1_sym_from_Lsg} is due the fact that the underestimation of $G_1(Q^2)$ by $\Ls(3Q^2/2)$, discussed in detail in Subsection~\ref{subsec:planardeg}, effectively captures the negative contribution to ${\overline \Gamma}_1^{\,\srm{sym}}(Q^2)$, furnished by the term $- G_3(Q^2)+G_4(Q^2)/4$ in \1eq{Gamma_1_sym_from_Gi}; recall, from \fig{fig:G1234}, that $G_4$ is mostly negative. 

This observation suggests that, at least in the symmetric limit,  
\1eq{meq} approximates the full 
$\overline{\fatg}^{\alpha\mu\nu}(q,r,p)$ 
even more accurately than the $\Ls(s^2)$ approximates the $G_1(q^2,r^2,p^2)$, by effectively capturing part of the contribution of the subleading form factors.

Finally, on the right panel of \fig{fig:Gamma_1_sym} we show the lattice result for ${\overline \Gamma}_1^{\,\srm{sym}}(Q^2)$ from~\mbox{\cite{Aguilar:2021lke,Boucaud:2002fx,Boucaud:2003xi}} (points). Note that these data are normalized in the so-called ``symmetric MOM scheme'', defined by the condition ${\overline \Gamma}_1^{\,\srm{sym}}(\mu^2) = 1$, with $\mu = 4.3$~GeV. Then, on the same panel, we show as a blue continuous line the result of \1eq{Gamma_1_sym_from_Lsg} after renormalization in the symmetric scheme\footnote{This is achieved by dividing the asymmetric scheme result for $\Ls(3Q^2/2)$ by its value at $Q^2 = \mu^2$.}. Clearly, \1eq{Gamma_1_sym_from_Lsg} approximates the lattice results quite well, with an error of less than $10\%$ for most of the range, which increases to $22\%$ at $r = 0.6$~GeV. As is evident from the agreement between ${\overline \Gamma}_1^{\,\srm{sym}}(Q^2)$ and $\Ls(3Q^2/2)$ on the left panel, the error with respect to the lattice originates from the truncation of the SDE, rather than from the use of the compact approximation given by \1eq{meq}.

\section{Conclusions}
\label{conc}

We have presented an extensive study of the transversely-projected three-gluon vertex 
by means of the SDE that determines its momentum evolution, 
making ample use of dynamical ingredients obtained from 
large-volume lattice simulations. 
The focal point of this investigation is the notable  
property of planar degeneracy: 
after a judicious choice of the tensor basis, 
the classical form factor of the vertex  
depends predominantly on a single kinematic variable, 
which represents a plane
in the space spanned by $q^2$, $r^2$, and $p^2$.
Our SDE-based approach
affords a valuable vantage point on the 
technical details surrounding this
special property, establishing 
its range of validity and degree of accuracy,  
for a particularly wide range of kinematic configurations.  
In fact, our analysis reveals that 
the planar degeneracy persists to a high degree of accuracy 
for general kinematics, and in particular 
for configurations that deviate 
completely from the bisectoral limit
($p^2 = r^2$), considered in~\cite{Pinto-Gomez:2022brg}.

Of central importance in the present study is the simple equation 
given by \1eq{meq}, which leads to a serious reduction of the 
technical effort required 
when dealing with the three-gluon vertex in nonperturbative  computations.
The relation given in \1eq{meq} emerges by combining the planar 
degeneracy with the observation that the classical form factor is considerably larger than all others.
The component $\Ls$ appearing in 
\1eq{meq} corresponds to the form factor of the soft-gluon limit,  a well-known quantity from a variety of lattice studies, and serves as a ``benchmark'' for the veracity of the SDE results.

The SDE analysis carried out probes the validity of  \1eq{meq},
and uses it in order to obtain a plethora of related results, which, in 
turn, demarcate its applicability. 
Our findings confirm 
that the classical form factor clearly dominates over the 
other three for a wide range of kinematics, 
with the exception of the region approaching the 
soft-gluon limit. However, due to the vanishing of the 
corresponding basis elements in this limit, 
\1eq{meq} represents 
an excellent approximation for all momenta.

One may wonder whether there exists a basis in which the 
planar degeneracy becomes exact. To be sure, it is always possible to construct such a basis, by absorbing the residual dependence of the ${\gammanew}_i(q^2,r^2,p^2)$ on the other two variables into the basis elements themselves, through
\begin{align}
\overline{\fatg}^{\,\alpha \mu \nu}(q,r,p) = \sum_{i=1}^4 {\widetilde G}_i(s^2) \, {\widetilde \tau}_{i}^{\alpha\mu\nu}(q,r,p)\,,  
\end{align}
with 
\be 
{\widetilde \tau}_{i}^{\alpha\mu\nu}(q,r,p) := {\gammanew}_i(q^2,r^2,p^2) \,
{\widetilde G}^{-1}(s^2) \, \tau_1^{\alpha \mu \nu}(q,r,p) \,.
\ee
Nonetheless, such a basis would be of no practical advantage,  
since it can only be constructed ``{\it a-posteriori}'',  namely 
after the form factors have been exactly determined. 
The truly remarkable observation is that there exists a ``simple'' basis containing the classical tensor as an element, which exhibits, in a natural way,
a rather accurate manifestation of planar degeneracy.

As mentioned in Sec.~\ref{sec:SDE3g},
the SDE for the three-gluon vertex
that we use is obtained from the 3PI three-loop effective action. The appropriate variation
of this action yields also 
the SDEs of the gluon and ghost propagator, as well  
and the SDE of the ghost-gluon vertex. 
All these equations are dynamically coupled to each other, 
and, from a strictly 
SDE-based point of view, they must be solved simultaneously,
as a coupled system comprised by numerous integral 
equations. 
Instead, in our approach we treat the 
SDE of the three-gluon vertex in isolation, 
using lattice ingredients for the gluon and ghost propagators.
Evidently, what one is tacitly assuming when adopting this 
approach 
is that the lattice results ``solve'' the corresponding SDEs 
to a very good approximation; that this is indeed so 
has been shown in detail in~\cite{Williams:2015cvx}, 
by solving the SDEs and comparing the solutions 
with the lattice results. 
We have therefore used the coincidence between 3PI SDEs and lattice, 
found in~\cite{Williams:2015cvx}, 
as our basic working hypothesis.

Finally, it would be interesting to investigate whether 
some generalized form of planar degeneracy holds for the 
transversely-projected four-gluon vertex~\cite{Pascual:1980yu,Papavassiliou:1992ia,Hashimoto:1994ct,Driesen:1998xc,Kellermann:2008iw,Ahmadiniaz:2013rla,Gracey:2014ola,Cyrol:2014kca,Binosi:2014kka,Eichmann:2015nra,Gracey:2017yfi}, which, just as the three-gluon vertex, 
is fully Bose-symmetric.  
Such a study may be particularly timely, 
given that lattice simulations are commencing to probe the basic structures of this vertex~\cite{Catumba:2021qbh}.

\section{Acknowledgments}
\label{sec:acknowledgments}

The work of  A.~C.~A. and L.~R.~S. are supported by the CNPq grants \mbox{307854/2019-1} and 
\mbox{162264/2022-4}.
A.~C.~A also acknowledges financial support from  project 464898/2014-5 (INCT-FNA).
M.~N.~F. and J.~P. are supported by the Spanish MICINN grant PID2020-113334GB-I00. M.~N.~F. acknowledges financial support from Generalitat Valenciana through contract \mbox{CIAPOS/2021/74}. J.~P. also acknowledges  
funding from the regional Prometeo/2019/087 from the Generalitat Valenciana.

\appendix

\section{Infrared divergence of the three-gluon vertex} 
\label{App_divergence} 

In this Appendix, we derive \1eq{asympt} for the asymptotic behavior of $\Ls(r^2)$ near the origin. Since the one-loop dressed gluonic diagrams are found to be infrared finite (see \fig{fig:lograte}), we focus on the contribution to $\Ls(r^2)$ originating from the ghost loops.

Starting with the expression for $d_{2,f}^{\,\scriptscriptstyle{sg}}(x,\mu^2)$ of \2eqs{lsg_renorm}{variosL}, we expand the term $F(u)B_1^2(u,y,\chi)$ appearing in $d_2^{\,\scriptscriptstyle{sg}}(x)$ around $x=0$. This procedure yields
\be 
d_{2,f}^{\,\scriptscriptstyle{sg}}(x,\mu^2) = - \frac{\lambda'}{3}\int_{0}^{\infty}\!\!\!\!dy\, f(y) \sqrt{\frac{y}{x}}\int_{0}^{\pi}\!\!\!\!d\phi\,  \frac{s_\phi^4c_\phi}{u}  - d_2^{\,\scriptscriptstyle{sg}}(\mu^2) + \ldots \,,
\ee
where $f(y) := F^3(y)B_1^3(y,y,\pi)$,
while the ellipsis denotes terms that are of higher order in $x$ and cannot contribute to the infrared divergence. Moreover, we used the fact that, in the soft-gluon limit, $B_1$ becomes independent of the angle.

Next, using the angular integral
\be
\sqrt{\frac{y}{x}}\int_{0}^{\pi}\!\!\!\!d\phi\,  \frac{s_\phi^4c_\phi}{u} = - \frac{\pi}{16x}\left[ \frac{y}{x^2}(y-2x)\theta( x - y ) + \frac{ x}{y^2}(x - 2 y )\theta(y - x )\right] \,, \label{angint}
\ee
where $\theta(x)$ denotes the Heaviside step function, we find
\be 
d_{2,f}^{\,\scriptscriptstyle{sg}}(x,\mu^2) = \frac{\lambda'\pi}{48}\left[ \int_{x}^{\infty}\!\!dy f(y)\frac{(x - 2 y)}{y^2} + \int_0^x\!\!\!\!dy\, f(y) \frac{y\left( y - 2x \right)}{x^3} \right] - d_2^{\,\scriptscriptstyle{sg}}(\mu^2) \,. \label{Ldiv_step1}
\ee

At this point, we recall the well-known anomalous dimensions of the ghost propagator and ghost-gluon vertex in Landau gauge, $F(y)\sim\ln^{-9/44}(y/\mu^2)$ and $B_1(y,y,\pi)\sim1$, respectively~\cite{vonSmekal:1997ern,Fischer:2002eq,Pennington:2011xs,Huber:2018ned}, for \mbox{$y\to \infty$}. Then, it follows that the first integral in \1eq{Ldiv_step1} contains an ultraviolet divergence, which ought to be canceled by the term $d_2^{\,\scriptscriptstyle{sg}}(\mu^2)$. Since the divergent part of $d_2^{\,\scriptscriptstyle{sg}}(\mu^2)$ results from the large $y$ behavior of its defining integral, it can be isolated by expanding the term $F(u)B_1^2(u,y,\chi)$ around $y \to \infty$, with the substitution $u \to \mu^2 + y - 2\mu\sqrt{y}c_\phi$. This expansion, together with another use of \1eq{angint}, results in  
\be 
d_{2}^{\,\scriptscriptstyle{sg}}(\mu^2) = \frac{\lambda'\pi}{48}\left[ \int_{\mu^2}^{\infty}\!\!dy f(y)\frac{(\mu^2 - 2 y)}{y^2} + \int_0^{\mu^2}\!\!\!\!dy\, f(y) \frac{y\left( y - 2\mu^2 \right)}{\mu^6} \right] \,,
\ee
up to finite terms. Hence, \1eq{Ldiv_step1} can be recast, after some algebra, as
\begin{align}
d_{2,f}^{\,\scriptscriptstyle{sg}}(x,\mu^2) =&\, \frac{\lambda'\pi}{48}\left[ \int_{x}^{\mu^2}\!\!dy f(y)\frac{(x - 2 y)}{y^2}  + (x - \mu^2) \int_{\mu^2}^\infty\!\!dy\frac{f(y)}{y^2}\, \right. \nonumber\\
+& \int_0^x\!\!\!\!dy\, f(y) \frac{y\left( y - 2x \right)}{x^3} - \left. \int_0^{\mu^2}\!\!\!\!dy\, f(y) \frac{y\left( y - 2\mu^2 \right)}{\mu^6} \right] \,,  \label{Ldiv_step2}
\end{align}
which is clearly ultraviolet finite.

The $x$-dependent integral in the second line of \1eq{Ldiv_step2} can be shown to be finite at $x = 0$, by means of an additional Taylor expansion. Then, the only possible divergence in \1eq{Ldiv_step2} appears 
when taking the limit $x\to 0$
of the first integral. Dropping finite terms, and integrating by parts, using $1/y = d\ln(y/\mu^2)/dy$ and \mbox{$1/y^2 = - d(1/y)/dy$}, one finds 
\begin{align} 
d_{2,f}^{\,\scriptscriptstyle{sg}}(x,\mu^2) =&\, \frac{\lambda'\pi}{48}\left\lbrace - 2 \left[ f(y) \ln \left(\frac{y}{\mu^2}\right)\right]_{x}
^{\mu^2} - x \left[ \frac{f(y)}{y} \right]_{x}
^{\mu^2} + x \left[ f^\prime(y) \ln\left(\frac{y}{\mu^2}\right) \right]_{x}
^{\mu^2} \right. \nonumber\\
+& \left. \int_x^{\mu^2}\!\!\!\!dy\, \ln \left(\frac{y}{\mu^2}\right)\left[ 2 f^\prime(y) - x f^{\prime\prime}(y)\right] \right\rbrace \,, \label{Ldiv_step3}
\end{align}
where primes denote derivatives with respect to $y$.

Now, the divergent integrand in \1eq{Ldiv_step3} is integrable at $x = 0$, and the only infrared divergence is in the lower limit of the first surface term. Hence, we have 
\be 
\lim_{x\to 0} \Ls(x) = \lim_{x\to 0}d_{2,f}^{\,\scriptscriptstyle{sg}}(x,\mu^2) = \frac{\alpha_s C_{\rm A}}{96\pi} F^3(0)B_1^3(0,0,\pi) \ln \left(\frac{x}{\mu^2}\right) \,, \label{Ldiv_step4}
\ee
up to finite terms.

To complete the derivation of \1eq{asympt}, let us recall that, in the Landau gauge, the quantum corrections to the ghost-gluon vertex are proportional to the ghost and anti-ghost momenta~\cite{Taylor:1971ff,Marciano:1977su,Aguilar:2018csq}, \ie
\be 
B_1(r^2,p^2,\theta_{rp}) = {\widetilde Z}_1 + r^\alpha p^\beta K_{\alpha\beta}(r,p,q) \,,
\ee
where $r\cdot p = r p \cos \theta_{rp}$, for some integral $K_{\alpha\beta}(r,p,q)$. In perturbation theory $K_{\alpha\beta}(r,p,q)$ can be shown to have pole divergences at $r = p = 0$, such that the corresponding limit of $B_1(r^2,p^2,\theta_{rp})$ is finite, but path-dependent. These divergences are eliminated by the mass of the gluon propagator. Thus, nonperturbatively
\be 
B_1(0,0,\theta_{rp}) = {\widetilde Z}_1 \,,
\ee
for any $\theta_{rp}$. Then, substituting the above result into \1eq{Ldiv_step4} we obtain  \1eq{asympt}.

%

\end{document}